\documentclass[11pt,draftcls,onecolumn]{IEEEtran}
\usepackage{graphicx}
\usepackage[cmex10]{amsmath}
\usepackage{amsfonts}
\usepackage{amssymb}
\usepackage{algorithmic}
\usepackage{algorithm}
\usepackage{array}
\usepackage{mdwmath}
\usepackage{mdwtab}
\usepackage[caption=false,font=footnotesize]{subfig}
\usepackage{fixltx2e}
\usepackage{stfloats}
\usepackage{footnote}
\usepackage{tabularx}
\usepackage{multirow}
\usepackage{color}

\newtheorem{definition}{Definition}

\newtheorem{theorem}{Theorem}

\newtheorem{remark}{Remark}

\begin{document}

\title{Intervention in Power Control Games \\With Selfish Users}

\author{Yuanzhang~Xiao,
        Jaeok~Park,
        and~Mihaela~van~der~Schaar,~\IEEEmembership{Fellow,~IEEE}
\thanks{Manuscript received March 31, 2011; revised August 15, 2011; accepted November 08, 2011.}
\thanks{Y. Xiao and M. van der Schaar are with the Electrical Engineering Department, University of California, Los Angeles, CA 90095 USA (e-mail: yxiao@ee.ucla.edu; mihaela@ee.ucla.edu).}
\thanks{J. Park is with the School of Economics, Yonsei University, Seoul 120-749, Korea (e-mail: jaeok.park@yonsei.ac.kr). He was previously with the Electrical Engineering Department, University of California, Los Angeles, CA 90095 USA.}
}

\maketitle

\begin{abstract}
We study the power control problem in single-hop wireless ad hoc networks with selfish users. Without incentive schemes, selfish users tend to
transmit at their maximum power levels, causing excessive interference to each other. In this paper, we study a class of incentive schemes based on
intervention to induce selfish users to transmit at desired power levels. In a power control scenario, an intervention scheme can be implemented by
introducing an intervention device that can monitor the power levels of users and then transmit power to cause interference to users if necessary.
Focusing on first-order intervention rules based on individual transmit powers, we derive conditions on the intervention rates and the power budget
to achieve a desired outcome as a (unique) Nash equilibrium with intervention and propose a dynamic adjustment process to guide users and the
intervention device to the desired outcome. We also analyze the effect of using aggregate receive power instead of individual transmit powers. Our
results show that intervention schemes can be designed to achieve any positive power profile while using interference from the intervention device
only as a threat. Lastly, simulation results are presented to illustrate the performance improvement from using intervention schemes and the
theoretical results.
\end{abstract}


\section{Introduction} \label{sec:intro}

Power control is an essential resource allocation scheme to control signal-to-interference-and-noise ratios (SINRs) for efficient transmission in
wireless networks. Extensive studies have been done on power control (see \cite{Chiang_FoundationTrend08} and references therein for an overview of
the literature in this topic). In many earlier works on power control, each user has a fixed minimum SINR requirement and then minimizes its transmit
power subject to the SINR requirement \cite[Ch. 2]{Chiang_FoundationTrend08}\cite{Yates95}\cite{Altman03}. This approach is suitable for fixed-rate
communications with voice applications. However, with the growing importance of data and multimedia applications, users are no longer satisfied with
a fixed SINR requirement, but they seek to maximize their utility reflecting the quality of service (QoS). To this end, most recent works formulate
the problem in the network utility maximization framework. In this framework, a central controller can compute optimal transmit power levels when the
utility functions are such that the network utility maximization problem is convex, and then assigns the optimal power levels to users. Assuming that
users are \emph{obedient} to the central controller, the problem can also be solved in a distributed manner \cite[Ch.
4]{Chiang_FoundationTrend08}\cite{HuangBerry_JSAC06}\cite{HandeChiang_ToN08}.

Besides the network utility maximization framework, many works use noncooperative games to model the distributed power control problem, in which each
user maximizes its own utility, instead of maximizing the network utility. In a noncooperative game model with a single frequency channel, each user
tends to transmit at its maximum power level to obtain high throughput, causing excessive interference to other users. This outcome may be far from
the global optimality of social welfare \cite{Chiang_FoundationTrend08}\cite{HuangBerry_JSAC06}\cite{Altman_Survey06}, especially when interference
among users is strong \cite{LarssonJorswieck09}. To improve the noncooperative outcome, various power control schemes have been proposed based on
pricing \cite{AlpcanBasar_CDMA}--\cite{CandoganOzdaglar_INFOCOM10}, auctions \cite{HuangBerry_06_Auction}, and mechanism design
\cite{SharmaTeneketzis_ToN}\cite{SharmaTeneketzis_GameTheory}. These works aim to achieve a better outcome by modifying the objective functions of
users using taxation and developing a distributed method based on the optimization of the modified objective functions. Users are assumed to be
obedient in that they accept the objective functions and follow the rule prescribed by the designer, and prices are used as control signals to guide
users to a desired outcome. However, \emph{selfish} users may have their own innate objectives which are different from the assigned objectives and
may ignore control signals and deviate from the prescribed rule if they are better off by doing so.

In summary, the methods in most existing works are not suitable for power control with selfish users. Selfish behavior of users can arise in many
practical scenarios without central controllers, such as wireless ad hoc networks, where each user transmits information from its own transmitter to
its own receiver, and multi-cell cellular networks, where the base station cannot control the interfering mobile stations in other cells. Hence, it
is important to design an incentive scheme to induce selfish users to achieve a desirable outcome in power control scenarios. One method to provide
incentives for selfish users is to impose taxation as real money payment. However, in order to achieve a desired outcome with a pricing scheme, the
designer needs to know how payment affects the payoffs of users, which is often the private information of users. The designer may use a mechanism
design approach as in \cite{SharmaTeneketzis_ToN}\cite{SharmaTeneketzis_GameTheory} to elicit private information, but it generally requires heavy
communication overheads.\footnote{Another drawback of \cite{SharmaTeneketzis_ToN} and \cite{SharmaTeneketzis_GameTheory} is the assumption that each
user's utility function is jointly concave in all the users' power levels, which seems to be unrealistic in power control scenarios.} Another method
to provide incentives is to use repeated games \cite{WuLiu09}\cite{LongZhang_JSAC07}. However, effective incentive schemes based on repeated games
require users to have long-run frequent interactions and to be sufficiently patient \cite{mailath}.

Recently, a new class of incentive schemes has been proposed based on the idea of intervention \cite{ParkMihaela_EURASIP}--\cite{ParkMihaela_JSAC}.
To implement an intervention scheme, we need an \emph{intervention device} that can monitor the actions of users and intervene in their interaction
if necessary. The \emph{monitoring technology} of the intervention device determines what it can observe about the actions of users, while its
\emph{intervention capability} determines the extent to which it can intervene in their interaction. An \emph{intervention rule} prescribes the
action that the intervention device should take as a function of its observation. Among existing works on intervention schemes,
\cite{ParkMihaela_EURASIP} and \cite{ParkMihaela_Gamenets} applied intervention schemes to contention games in the medium access control (MAC) layer,
while \cite{ParkMihaela_JSAC} studied the impact of the monitoring technology and the intervention capability on the system performance in an
abstract model. We also note that \cite{GaiKrishnamachari_Infocom} proposed a packet-dropping mechanism for queueing games using an idea similar to
intervention. In this paper, we focus on a power control scenario and study intervention schemes in this particular scenario.

In the power control scenario considered in this paper, the intervention device estimates the individual transmit power of each user or the aggregate
receive power at its receiver and then transmits at a certain power level following the intervention rule prescribed by the designer. In order to
achieve a target operating point, the designer can use an intervention rule such that the intervention device transmits minimum, possibly zero, power
if users are transmitting at the desired power levels, while transmitting at a high power level to reduce the SINRs of users if a deviation is
detected. In this way, an intervention scheme can punish the misbehavior of users and regulate the power transmission of selfish users. We first
consider a monitoring technology with which the intervention device can estimate the individual transmit power of each user without errors. While
focusing on a simple class of intervention rules called first-order intervention rules, we study the requirements for the parameters of first-order
intervention rules to achieve a given target power profile as a (unique) Nash equilibrium (NE). We propose a dynamic adjustment process that the
designer can use to guide users to the target power profile through intermediate targets. We then relax the monitoring requirement and consider a
monitoring technology with which the intervention device can estimate only the aggregate receive power. We show that with aggregate observation,
intervention rules can be designed to achieve a given target as a NE but rarely as a unique NE. Our results provide a systematic design principle
based on which a designer can choose an intervention scheme (an intervention device and an intervention rule) to achieve a desired outcome. Our
analysis suggests that, unlike pricing schemes, it is possible for the designer to design effective intervention schemes without having knowledge
about how users value their SINRs, as long as their utility is monotonically increasing with their own SINRs. We also propose a method based on
intervention for the designer to estimate the cross channel gains, the noise powers, and maximum transmit power levels of users without any
cooperative behavior of users such as sending pilot signals for channel estimation and reporting the estimates to the designer. After obtaining
relevant information, the designer can configure an intervention rule to achieve a target operating point as NE.

The rest of the paper is organized as follows. In Section~\ref{sec:model}, we describe the system model of power control with intervention. In
Section~\ref{sec:PerformanceMetrics}, we propose design criteria for intervention rules, performance characteristics to evaluate intervention rules,
and classes of intervention rules. In Section~\ref{sec:firstorder_perfectmonitoring}, we study the design of first-order intervention rules to
achieve a target power profile. In Section~\ref{sec:implementation}, we discuss implementation issues related to intervention. Simulation results are
presented in Section~\ref{sec:Simulation}. Finally, Section~\ref{sec:conclusion} concludes the paper. For the ease of reference, we summarize major
notation used in this paper in Table~\ref{table:Notation}.

\begin{table}
\renewcommand{\arraystretch}{1.1}
\caption{Summary of Major Notation.} \label{table:Notation} \centering
{
\begin{tabular}{|c|l|}
\hline Symbol & Description \\
\hline $N$ & Number of (regular) users in the network\\
\hline $\mathcal{N}$ & Set of users, $\mathcal{N}\triangleq\{1,\ldots,N\}$ \\
\hline $P_i$ & Maximum transmit power level of user $i$; $i \in \mathcal{N}$ \\
\hline $P_0$ & Power budget of the intervention device, or the intervention capability \\
\hline $\mathcal{P}_i$ & User $i$'s action space $\mathcal{P}_i\triangleq[0,P_i]$; $i\in\mathcal{N}\cup\{0\}$ \\
\hline $p_i$ & Transmit power level of user $i$, $p_i \in \mathcal{P}_i$; $i \in \mathcal{N}\cup\{0\}$ \\
\hline $\mathbf{p}$ & Power profile of the users, $\mathbf{p}=(p_1,\ldots,p_N) \in \mathcal{P}
\triangleq \prod_{i=1}^N \mathcal{P}_i$ \\
\hline $\mathbf{p}^\star$ & Target power profile, $\mathbf{p}^\star=(p_1^\star,\ldots,p_N^\star)$ \\
\hline $\mathbf{P}$ & Profile of the maximum power levels, $\mathbf{P}=(P_1,\ldots,P_N)$ \\
\hline $h_{ij}$ & Channel gain from user $j$'s transmitter to user $i$'s receiver; $i,j\in\mathcal{N}\cup\{0\}$ \\
\hline $n_i$ & Noise power at user $i$'s receiver \\
\hline $\gamma_i$ & SINR at user $i$'s receiver \\
\hline $\mathcal{Y}$ & Set of all possible signals obtained by the intervention device \\
\hline $\rho$ & Signal determination function, $\rho:\mathcal{P} \rightarrow \mathcal{Y}$ \\
\hline $f$ & Intervention rule, $f:\mathcal{Y}\rightarrow \mathcal{P}_0$ \\
\hline $\mathcal{E}(f)$ & Set of power profiles sustained by $f$ \\
\hline $\mathcal{F}_{K}(\mathbf{p}^\star)$ & Set of $K$th-order intervention rules with target power profile $\mathbf{p}^\star$ \\
\hline $PB_K(\mathbf{p}^\star)$ & Minimum power budget to sustain $\mathbf{p}^\star$ using $K$th-order intervention rules \\
\hline $PB_K^s(\mathbf{p}^\star)$ & Minimum power budget to strongly sustain $\mathbf{p}^\star$ using $K$th-order intervention rules \\
\hline $f_1^I$ & First-order intervention rule based on individual transmit powers \\
\hline $\alpha_i$ & Intervention rate for user $i$ in $f_1^I$; $i\in\mathcal{N}$ \\
\hline $f_1^A$ & First-order intervention rule based on aggregate receive power \\
\hline $\alpha_0$ & Aggregate intervention rate in $f_1^A$ \\
\hline $\mathcal{\tilde{N}}(\mathbf{p}^\star)$ & Set of users whose target powers are less than the maximum power levels,
$\mathcal{\tilde{N}}(\mathbf{p}^{\star}) = \{ i \in \mathcal{N}:
p_i^{\star} < P_i \}$ \\
\hline $N'$ & Number of users whose target powers are less than the maximum power levels, $N'=|\mathcal{\tilde{N}}(\mathbf{p}^\star)|$ \\
\hline $\{(f^t,\mathbf{p}^t)\}_{t=1}^T$ & Sequence of intervention rules and power profiles in the dynamic adjustment process \\
\hline $\{\mathbf{\tilde{p}}^t\}_{t=1}^T$ & Sequence of intermediate target power profiles in the dynamic adjustment process \\
\hline $d(\mathbf{p},\mathbf{p}')$ & Relative distance between two power profiles $\mathbf{p}$ and $\mathbf{p}'$, $d(\mathbf{p},\mathbf{p}') = \sum_{i=1}^N (p_i - p'_i)/p_i$ \\
\hline $T^\star(\mathbf{p}^\star)$ & Minimum convergence time (in steps) for the dynamic adjustment process to reach $\mathbf{p}^\star$ \\
\hline
\end{tabular}
}
\end{table}

\section{Model of Power Control With Intervention} \label{sec:model}

We consider a single-hop wireless ad hoc network, where a fixed set of $N$ users and an intervention device transmit in a single frequency channel.
The set of the users is denoted by $\mathcal{N} \triangleq \{1,2,\ldots,N\}$. Each user has a transmitter and a receiver. Each user $i$ chooses its
transmit power $p_i$ in the set $\mathcal{P}_i \triangleq [0,P_i]$, where $P_i > 0$ for all $i \in \mathcal{N}$. The power profile of all the users
is denoted by ${\bf p}=(p_1,\ldots,p_N) \in \mathcal{P} \triangleq \prod_{i=1}^N \mathcal{P}_i$, and the power profile of all the users other than
user $i$ is denoted by ${\bf p}_{-i}$.

In the network, there is an intervention device, sometimes referred to as user $0$, that consists of a transmitter and a receiver. The receiver of
the intervention device can monitor the power profile of the users, while the transmitter can create interference to the users by transmitting power.
Once the users choose their power profile, the intervention device obtains a signal $y \in \mathcal{Y}$, where $\mathcal{Y}$ is the set of all
possible signals. We assume that the signal is realized deterministically given a power profile and use the signal determination function $\rho:
\mathcal{P}\rightarrow \mathcal{Y}$ to denote the signal given the power profile $\mathbf{p}$.\footnote{More generally, we can assume that the signal
is realized randomly given a power profile and use $\rho(\mathbf{p})$ to represent the probability distribution of signals given $\mathbf{p}$. We
leave the analysis of this general case for future research.} After observing a signal, the intervention device chooses its own transmit power $p_0$
in the set $\mathcal{P}_0 \triangleq [0,P_0]$, where $P_0>0$. We call $(\mathcal{Y}, \rho)$ the monitoring technology of the intervention device, and
call $P_0$ its intervention capability. The ability of an intervention device is characterized by its monitoring technology and intervention
capability. We call the transmit powers of the intervention device and the users $(p_0, \mathbf{p}) \in \mathcal{P}_0 \times \mathcal{P}$ an outcome.

The QoS obtained by user $i$ is determined by its SINR, denoted by $\gamma_i$. We use a block-fading channel model with sufficiently long fading
blocks, as in \cite{Yates95}--\cite{HandeChiang_ToN08},\cite{LarssonJorswieck09}--\cite{LongZhang_JSAC07}. In one block, for $i,j \in \mathcal{N}
\cup \{0\}$, let $h_{ij}$ be the channel gain from user $j$'s transmitter to user $i$'s receiver, and let $n_i$ be the noise power at user $i$'s
receiver. If code-division multiple accessing (CDMA) is used, the channel gain is defined as the effective channel gain taking into account the
spreading factor. In one block, the users and the intervention device transmit at constant power levels $\mathbf{p}$ and $p_0$,
respectively.\footnote{In practice, there is a time lag between when the users transmit and when the intervention device transmits because the
intervention has to monitor the users' power profile in order to decide its transmit power level. In this paper, we assume that this time lag is
short and negligible compared to the length of fading blocks, although in principle we can take into account the time lag in the users' utility
functions. See, for example, \cite{ParkMihaela_Gamenets} for a model that takes into account the time lag explicitly.} Then the SINR of user $i \in
\mathcal{N}$ is given by
\begin{eqnarray}
\gamma_i(p_0,{\bf p}) = \frac{h_{ii} p_i}{h_{i0} p_0 + \sum_{j\neq i} h_{ij} p_j + n_i}.\footnote{}
\end{eqnarray}
\footnotetext{Throughout the paper, we use $j \neq i$ with the summation operator to mean $j \in \mathcal{N} \setminus \{i\}$, not $j \in \mathcal{N}
\cup \{0\} \setminus \{i\}$.} We assume that each user $i \in \mathcal{N}$ has monotonic preferences on its own SINR in the sense that it weakly
prefers $\gamma_i$ to $\gamma'_i$ if and only if $\gamma_i \geq \gamma'_i$. Our analysis does not require any other properties of preferences (for
example, preferences do not need to be represented by a concave utility function).\footnote{The preferences of a user may be defined on dimensions
other than its SINR. For example, a user may use the ratio of SINR to transmit power as the utility function, as in \cite{SaraydarMandayam_TCOM02},
because it may care about its energy consumption as well. Our approach can be applied to such a scenario if utility functions representing the users'
preferences are known to the designer.}

In our setting, the intervention device has a receiver to measure the aggregate receive power from all the users. Furthermore, if the receiver moves
and takes measurement at different locations, it can estimate the individual transmit power of each user as well. Thus, in this paper we will focus
on two types of monitoring technology with which the intervention device can estimate individual transmit powers $\mathbf{p}$ or an aggregate receive
power $\sum_{i=1}^N h_{0i} p_i$. { In other words, we consider two signal determination functions, $\rho(\mathbf{p})=\mathbf{p}$ and
$\rho(\mathbf{p})=\sum_{i=1}^N h_{0i} p_i$.}

\section{Design of Intervention Rules}\label{sec:PerformanceMetrics}

\subsection{Design Criteria}

Since the intervention device transmits its power after it obtains a signal, its strategy can be represented by a mapping $f: \mathcal{Y} \rightarrow
\mathcal{P}_0$, which is called an intervention rule. The SINR of user $i$ when the intervention device uses an intervention rule $f$ and the users
choose a power profile $\mathbf{p}$ is given by $\gamma_i(f(\rho(\mathbf{p})),{\bf p})$. With an abuse of notation, we will use $f(\mathbf{p})$ to
mean $f(\rho(\mathbf{p}))$. Given an intervention rule $f$, the interaction among the users that choose their own power levels selfishly can be
modeled as a non-cooperative game, whose strategic form is given by
\begin{eqnarray}
\Gamma_{f} = \langle \mathcal{N},(\mathcal{P}_i)_{i \in \mathcal{N}}, (\gamma_i(f(\cdot),\cdot))_{i \in \mathcal{N}} \rangle.
\end{eqnarray}
We can predict the power profile chosen by the users given an intervention rule using the concept of Nash equilibrium.

\begin{definition}
A power profile $\mathbf{p}^* \in \mathcal{P}$ is a Nash equilibrium of the game $\Gamma_{f}$ if
\begin{eqnarray}
\gamma_i(f(\mathbf{p}^*),\mathbf{p}^*) \geq \gamma_i(f(p_i,\mathbf{p}_{-i}^*),p_i,\mathbf{p}_{-i}^*)
\end{eqnarray}
for all $p_i \in \mathcal{P}_i$ and all $i \in \mathcal{N}$.
\end{definition}

When a power profile $\mathbf{p}^*$ is a NE of $\Gamma_{f}$, no user has the incentive to deviate from $\mathbf{p}^*$ unilaterally provided that the
intervention device uses intervention rule $f$. Moreover, if $\mathbf{p}^*$ is a unique NE of $\Gamma_{f}$, intervention has robustness in that the
designer does not need to worry about coordination failure (i.e., the possibility that the users get stuck in a ``wrong'' equilibrium).

\begin{definition}
An intervention rule $f$ \emph{(strongly) sustains} a power profile $\mathbf{p}^*$ if $\mathbf{p}^*$ is a (unique) NE of the game $\Gamma_f$.
\end{definition}

We use $\mathcal{E}(f)$ to denote the set of all power profiles sustained by $f$. Suppose that the designer has a welfare function $U_0(\gamma_1,
\ldots, \gamma_N)$, defined on the users' SINRs. Then the objective of the designer is to find a target power profile that maximizes the welfare
function and an intervention rule that (strongly) sustains the target power profile. Formally, the design problem solved by the designer can be
written as
\begin{eqnarray} \label{eq:desprob}
\max_{\mathbf{p}} \max_{f} \{ U_0(\gamma_1(f(\mathbf{p}),\mathbf{p}), \ldots, \gamma_N(f(\mathbf{p}),\mathbf{p})) : \mathbf{p} \in \mathcal{E}(f) \}.
\end{eqnarray}
If uniqueness is desired, we can replace $\mathbf{p} \in \mathcal{E}(f)$ in \eqref{eq:desprob} with $\{\mathbf{p}\} = \mathcal{E}(f)$. Note that a
solution to the design problem \eqref{eq:desprob}, $\mathbf{p}^\star$ and $f^*$, must satisfy $f^*(\mathbf{p}^\star)=0$, namely no intervention if
the users choose the target power profile. Based on this observation, the design problem \eqref{eq:desprob} can be solved in two steps. In the first
step, we obtain a target power profile $\mathbf{p}^\star$ by solving
\begin{eqnarray}\label{eq:SolveForTargetPowerProfile}
\max_{\mathbf{p}} \{ U_0(\gamma_1(0,\mathbf{p}), \ldots, \gamma_N(0,\mathbf{p})) : \mathbf{p} \in \mathcal{P} \}.
\end{eqnarray}
There always exists a solution to the optimization problem \eqref{eq:SolveForTargetPowerProfile} as long as the welfare function $U_0$ is continuous
in $\gamma_1,\ldots,\gamma_N$. Under some welfare functions, e.g. $U_0(\gamma_1, \ldots, \gamma_N)\sum_{i=1}^N \log(\gamma_i)$ as in
\cite{HuangBerry_JSAC06}\cite{HandeChiang_ToN08}, the optimization problem \eqref{eq:SolveForTargetPowerProfile} is convex, and thus easy to solve.
If \eqref{eq:SolveForTargetPowerProfile} is solved off-line, the designer can choose other welfare functions even though the resulting optimization
problem is nonconvex. In the second step, we look for an intervention rule $f$ such that $f(\mathbf{p}^\star)=0$ and $\mathbf{p}^\star \in
\mathcal{E}(f)$ (or $\{\mathbf{p}^\star\} = \mathcal{E}(f)$), given the target power profile $\mathbf{p}^\star$ obtained in the first step.

The first step of solving the design problem \eqref{eq:desprob}, namely finding the target power profile $\mathbf{p}^\star$, requires knowledge about
the parameters in the model, $h_{ij}$, $n_i$, and $P_i$ for all $i,j \in \mathcal{N} \cup \{0\}$. In Section~\ref{sec:implementation}-A, we propose a
method for the designer to estimate the relevant system parameters needed to solve \eqref{eq:SolveForTargetPowerProfile} based on intervention. Note,
however, that solving the problem \eqref{eq:SolveForTargetPowerProfile} does not require the designer to know the details of the users' preferences
on their SINRs since knowing the expressions for SINRs suffices to evaluate the welfare function and to check the incentive constraints. To highlight
the informational advantage of our approach, let us consider a pricing scheme, in which each user~$i$ is charged $\pi_i(\mathbf{p})$ when the users
choose a power profile $\mathbf{p}$. Suppose that each user $i$ has quasilinear preferences on its own SINR and payment which are represented by a
utility function of the form $u_i(\gamma_i) - \pi_i$. Then in order to find a pricing scheme that sustains a target profile $\mathbf{p}^{\star}$, the
designer needs to know $u_i$ for all $i \in \mathcal{N}$. Since a pricing scheme uses an outside instrument to influence the decisions of the users,
the designer needs to know how the users value SINRs relative to payments, which is subjective and thus hard to measure. On the contrary, an
intervention scheme affects the users through their SINRs, and thus the designer does not need to know how the users value their SINRs.

In the subsequent discussion of this paper, we focus on the second step of solving the design problem \eqref{eq:desprob}, assuming that a target
power profile $\mathbf{p}^{\star}$ has been found. That is, we aim to find an intervention rule $f$ that (strongly) sustains $\mathbf{p}^\star$,
namely $\mathbf{p}^{\star} \in \mathcal{E}(f)$ (or $\{\mathbf{p}^{\star}\} = \mathcal{E}(f)$), and that satisfies $f(\mathbf{p}^{\star}) = 0$. Since
user $i$ can guarantee a positive SINR by choosing a positive power, it is impossible to provide an incentive for user $i$ to choose $p_i = 0$ using
any intervention rule. Thus, we assume that $\mathbf{p}^{\star} \in \prod_i (0,P_i]$. We say that $(f,\mathbf{p}^{\star})$ is an \emph{(intervention)
equilibrium} if $\mathbf{p}^{\star} \in \mathcal{E}(f)$ and $f(\mathbf{p}^{\star}) = 0$. At an equilibrium, no user has an incentive to deviate
unilaterally while the designer fulfills his design criteria. Thus, an equilibrium can be considered as a stable configuration of an intervention
rule and a power profile. An equilibrium can be achieved following the procedure described below.
\begin{enumerate}
\item The designer chooses a target power profile $\mathbf{p}^{\star}$ and an intervention rule $f$.
\item The users choose their power profile $\mathbf{p}$, knowing the intervention rule chosen by the designer.\footnote{The
intervention rule can be broadcast to the users, or learned by the users from experimentation.}
\item The intervention device obtains a signal $\rho(\mathbf{p})$ and chooses its power $p_0 = f(\mathbf{p})$.
\end{enumerate}
To execute the above procedure, we may consider an adjustment process (e.g., one based on best-response updates) for the users and the intervention
device to reach an equilibrium (see Section~\ref{sec:firstorder_perfectmonitoring}-B), as well as an estimation process for the intervention device
to obtain a signal (see Section~\ref{sec:implementation}-A). It is an implicit underlying assumption of our analysis that the time it takes to reach
a final outcome (i.e., the duration of the procedure) is short relative to the time for which the final outcome lasts. This justifies that in our
model, the users fully take into account interference from the intervention device that is realized at the final outcome when they make decisions
about their powers. When a network parameter changes (e.g., some users leave or join the network, or move to different locations), a new target is
chosen and the procedure is repeated to achieve a new equilibrium. Thus, our analysis holds as long as network parameters do not change frequently,
whereas providing incentives using a repeated game strategy usually requires an infinite horizon and sufficiently patient players.

Another important underlying assumption in our analysis is that the designer can commit to the intervention rule it chooses. Since $U_0$ is
increasing in each $\gamma_i$ and each $\gamma_i$ is decreasing in $p_0$, the designer prefers not to intervene at all, i.e., it prefers to choose
$p_0 = 0$. However, if $p_0$ is held fixed at $0$, the users will choose $\mathbf{P}$. The role of the intervention device is to provide a punishment
mechanism for the users to choose a desired power profile other than $\mathbf{P}$; the device should choose a high power level if a deviation is
detected. Without the designer's commitment to the intervention rule, the designer would choose $p_0 = 0$ (e.g., by disabling the intervention
device) regardless of the power profile chosen by the users. Predicting this behavior of the designer, the users would choose $\mathbf{P}$, resulting
in the same outcome as with no intervention. Therefore, in order for the proposed intervention schemes to provide incentives successfully, it is
critical that the designer executes punishment as promised to make punishment credible. In practice, credibility can be achieved by programming the
intervention rule in the intervention device and making the program difficult to manipulate.

Finally, the benchmark is the welfare when there is no intervention device in the network, i.e., $p_0$ is held fixed at $0$. In this case, $\gamma_i$
is always strictly increasing in $p_i$, and thus $\mathbf{P} \triangleq (P_1, \ldots, P_N)$ is the unique NE of the game $\langle
\mathcal{N},(\mathcal{P}_i)_{i \in \mathcal{N}}, (\gamma_i(0,\cdot))_{i \in \mathcal{N}} \rangle$.\footnote{This is true for any constant
intervention rule, where $p_0$ is chosen independently of the observation of the intervention device. This shows the inability of traditional
Stackelberg games, where the leader (the intervention device) takes an action before the followers (the users) do, to provide incentives in our
setting.} Thus, the case of $\mathbf{p}^{\star} = \mathbf{P}$ is trivial, because there is no need to use an incentive scheme in order to achieve the
power profile $\mathbf{P}$. Hence, our main interest lies in the case of $\mathbf{p}^{\star} \neq \mathbf{P}$, although our analysis does not exclude
the case of $\mathbf{p}^{\star} = \mathbf{P}$. The case of $\mathbf{p}^{\star} \neq \mathbf{P}$ arises in many situations, for example, when
interference among users is strong and welfare is measured by the sum of utilities or the minimum of utilities
\cite{Chiang_FoundationTrend08}\cite{HuangBerry_JSAC06}\cite{Altman_Survey06}\cite{LarssonJorswieck09}.

\subsection{Performance Characteristics}

Given a target power profile $\mathbf{p}^{\star}$, there are potentially many intervention rules $f$ that satisfy the design criteria
$\mathbf{p}^{\star} \in \mathcal{E}(f)$ and $f(\mathbf{p}^{\star}) = 0$. Thus, below we propose several performance characteristics with which we can
evaluate different intervention rules satisfying the design criteria.

1) Monitoring requirement: The minimum amount of information about the power profile that is required for the intervention device to execute a given
intervention rule (assuming perfect estimation).

2) Intervention capability requirement: The minimum intervention capability needed for the intervention device to execute a given intervention rule,
i.e., $\sup_{\mathbf{p} \in \mathcal{P}} f(\mathbf{p})$. (Even though there is no intervention at an equilibrium, the intervention device should have
an intervention capability $P_0 \geq \sup_{\mathbf{p} \in \mathcal{P}} f(\mathbf{p})$ in order to make the intervention rule $f$ credible to the
users.)

3) Strong sustainment: Whether a given intervention rule strongly sustains the target power profile $\mathbf{p}^{\star}$.

4) Complexity: The complexity of a given intervention rule in terms of design, broadcast/learning, and computation.

\subsection{Classes of Intervention Rules}

Without loss of generality, we can express an intervention rule $f$ satisfying $f(\mathbf{p}^{\star}) = 0$ as $f(\mathbf{p}) =
[g(\mathbf{p})]_0^{P_0}$, where $[x]_a^b = \min \{\max\{x,a\},b\}$, for some function $g: \mathcal{P} \rightarrow \mathbb{R}$ such that
$g(\mathbf{p}^{\star}) = 0$. Also, since the designer desires to achieve $\mathbf{p}^{\star}$, it is natural to consider functions $g$ that increase
as the users deviate from $\mathbf{p}^{\star}$. Hence, we consider the following classes of intervention rules,
\begin{eqnarray}
\mathcal{F}_{K}(\mathbf{p}^{\star}) = \left\{ f: f(\mathbf{p}) = \left[ \sum_{i=1}^N \sum_{k=1}^K \alpha_{i,k} |p_i - p_i^{\star}|^k \right]_0^{P_0}
\text{ for some $\alpha_{i,k} \geq 0$ and $P_0 > 0$} \right\},
\end{eqnarray}
for $K = 1, 2, \ldots$. We call an intervention rule $f \in \mathcal{F}_{K}(\mathbf{p}^{\star})$ a \emph{$K$th-order intervention rule with target
power profile $\mathbf{p}^{\star}$}. As $K$ becomes larger, the set $\mathcal{F}_{K}(\mathbf{p}^{\star})$ contains more intervention rules, but at
the same time complexity increases. Simple intervention rules are desirable for the designer, the users, and the intervention device. As $K$ is
smaller, the designer has less design parameters and less burden of broadcasting the intervention rule; the users can more easily learn the
intervention rule and find their best responses during an adjustment process; and the intervention device can more quickly compute the value of the
intervention rule at the chosen power profile. Thus, our analysis mainly focuses on first-order intervention rules, the simplest among the above
classes.

Let $\mathcal{\tilde{F}}_{K}(\mathbf{p}^{\star})$ ($\mathcal{\tilde{F}}_{K}^s(\mathbf{p}^{\star})$) be the set of all $K$th-order intervention rules
that (strongly) sustains $\mathbf{p}^{\star}$, i.e., $\mathcal{\tilde{F}}_{K}(\mathbf{p}^{\star}) = \{ f \in \mathcal{F}_{K}(\mathbf{p}^{\star}) :
\mathbf{p}^{\star} \in \mathcal{E}(f) \}$ and $\mathcal{\tilde{F}}_{K}^s(\mathbf{p}^{\star}) = \{ f \in \mathcal{F}_{K}(\mathbf{p}^{\star}) :
\{\mathbf{p}^{\star}\} = \mathcal{E}(f) \}$. We define the \emph{minimum power budget}{\footnote{As to be shown later, for strong sustainment, we
need $P_0$ to exceed a certain value. Thus, we actually mean ``infimum'' power budget by minimum power budget.}} for a $K$th-order intervention rule
to (strongly) sustain $\mathbf{p}^{\star}$ by
\begin{eqnarray}
PB_K(\mathbf{p}^{\star}) = \inf_{f \in \mathcal{\tilde{F}}_{K}(\mathbf{p}^{\star})} \sup_{\mathbf{p} \in \mathcal{P}}
f(\mathbf{p}),~\mathrm{and}~PB_K^s(\mathbf{p}^{\star}) = \inf_{f \in \mathcal{\tilde{F}}_{K}^s(\mathbf{p}^{\star})} \sup_{\mathbf{p} \in \mathcal{P}}
f(\mathbf{p}).
\end{eqnarray}
Thus, with an intervention capability $P_0 > PB_K(\mathbf{p}^{\star})$ ($P_0 > PB_K^s(\mathbf{p}^{\star})$), there exists a $K$th-order intervention
rule that (strongly) sustains $\mathbf{p}^{\star}$. We set $PB_K(\mathbf{p}^{\star}) = + \infty$ ($PB_K^s(\mathbf{p}^{\star}) = + \infty$) if there
is no $K$th-order intervention rule that (strongly) sustains $\mathbf{p}^{\star}$ (i.e., $\mathcal{\tilde{F}}_{K}(\mathbf{p}^{\star})$
($\mathcal{\tilde{F}}_{K}^s(\mathbf{p}^{\star})$) is empty). Since $\mathcal{F}_K(\mathbf{p}^{\star}) \subset \mathcal{F}_{K'}(\mathbf{p}^{\star})$
for all $K, K'$ such that $K \leq K'$, both $PB_K(\mathbf{p}^{\star})$ and $PB_K^s(\mathbf{p}^{\star})$ are weakly decreasing in $K$ for all
$\mathbf{p}^{\star}$. This suggests a trade-off between complexity and the minimum power budget. Also, since
$\mathcal{\tilde{F}}_{K}^s(\mathbf{p}^{\star}) \subset \mathcal{\tilde{F}}_{K}(\mathbf{p}^{\star})$, we have $PB_K(\mathbf{p}^{\star}) \leq
PB_K^s(\mathbf{p}^{\star})$ for all $K$ and $\mathbf{p}^{\star}$. The difference $PB_K^s(\mathbf{p}^{\star}) - PB_K(\mathbf{p}^{\star})$ can be
interpreted as the price of strong sustainment in terms of the minimum power budget.

\section{Analysis of First-Order Intervention Rules}\label{sec:firstorder_perfectmonitoring}
\subsection{Design of Intervention Rules}

We consider first-order intervention rules of the form
\begin{eqnarray}
f_1^I(\mathbf{p}) = \left[\sum_{i=1}^N \alpha_i |p_i - p_i^{\star}| \right]_0^{P_0}.
\end{eqnarray}
Under a first-order intervention rule, the intervention device increases its transmit power linearly with the deviation of each user from the target
power, $|p_i - p_i^{\star}|$, in the range of its intervention capability. We call $\alpha_i$ the \emph{intervention rate} for user $i$, which
measures how sensitive intervention reacts to a deviation of user $i$. Let $\mathcal{\tilde{N}}(\mathbf{p}^{\star}) = \{ i \in \mathcal{N}:
p_i^{\star} < P_i \}$. Without loss of generality, we label the users in such a way that $i \in \mathcal{\tilde{N}}(\mathbf{p}^{\star})$ if and only
if $i \leq N'$, where $N' = |\mathcal{\tilde{N}}(\mathbf{p}^{\star})|$. Since the users have natural incentives to choose their maximum powers in the
absence of intervention, we need to provide incentives only for the users in $\mathcal{\tilde{N}}(\mathbf{p}^{\star})$. The following theorem shows
that when the intervention capability is sufficiently large, the designer can always find intervention rates to have a given target power profile
$\mathbf{p}^{\star}$ sustained by a first-order intervention rule.

\begin{theorem}\label{theorem:FirstOrderStrict_CompleteInfo}
For any $\mathbf{p}^{\star} \in \prod_i (0,P_i]$, $\mathbf{p}^{\star} \in \mathcal{E}(f_1^I)$ if and only if
\begin{eqnarray} \label{eq:alphai}
\alpha_i \geq \frac{\sum_{j\neq i} h_{ij} p_j^{\star} + n_i}{p_i^{\star}h_{i0}}
\end{eqnarray}
and
\begin{eqnarray} \label{eq:P0firstorder}
P_0 \geq \frac{(P_i-p_i^{\star}) (\sum_{j\neq i} h_{ij} p_j^{\star} + n_i)}{p_i^{\star}h_{i0}}
\end{eqnarray}
for all $i \in \mathcal{\tilde{N}}(\mathbf{p}^{\star})$.
\end{theorem}

\begin{IEEEproof}
See \cite[Appendix~B]{XiaoParkMihaela_JSTSP}.
\end{IEEEproof}

We can interpret Theorem~\ref{theorem:FirstOrderStrict_CompleteInfo} as follows. As $h_{i0}$ is larger, intervention causes more interference to user
$i$ with the same transmit power, and thus the intervention rate $\alpha_i$ can be chosen smaller to yield the same interference. When $\sum_{j\neq
i} h_{ij} p_j^{\star} + n_i$ is large, interference to user $i$ from other users and its noise power are already strong, and thus the intervention
rate $\alpha_i$ should be large in order for intervention to be effective. Hence, $h_{i0}/(\sum_{j\neq i} h_{ij} p_j^{\star} + n_i)$ can be
interpreted as the effectiveness of intervention to user  $i$. Without intervention, users have natural incentives to increase their transmit powers.
Thus, as the target power for user $i$, $p_i^{\star}$, is smaller, the incentive for user $i$ to deviate is stronger, and thus a larger intervention
rate $\alpha_i$ is needed to prevent deviation. Note that $(P_i-p_i^{\star})$ is the maximum possible deviation by user $i$ (in the direction where
it has a natural incentive to deviate). The minimum intervention capability in the right-hand side of \eqref{eq:P0firstorder} is increasing with the
maximum possible deviation and the strength of the incentive to deviate while decreasing with the effectiveness of intervention.

A first-order intervention rule $f_1^I$ satisfying the conditions in Theorem~\ref{theorem:FirstOrderStrict_CompleteInfo} may have a NE other than the
target power profile $\mathbf{p}^{\star}$. For example, if $P_0 \leq \sum_{j \neq i} \alpha_j (P_j - p_j^{\star})$ for all $i \in
\mathcal{\tilde{N}}(\mathbf{p}^{\star})$, $\mathbf{P}$ is also sustained by $f_1^I$. The presence of this extra NE is undesirable since it brings a
possibility that the users still choose $\mathbf{P}$ while the intervention device causes interference to the users by transmitting its maximum power
$P_0$. Obviously, this outcome $(P_0, \mathbf{P})$ is worse for every user than the outcome at the unique NE without intervention $(0,\mathbf{P})$.
In order to eliminate this possibility, the designer may want to choose an intervention rule that strongly sustains the target power profile. The
following theorem provides a sufficient condition for a first-order intervention rule to strongly sustain a given target power profile.

\begin{theorem}\label{theorem:UniqueNE_FirstOrderStrict_CompleteInfo}
For any $\mathbf{p}^{\star} \in \prod_i (0,P_i]$, $\{\mathbf{p}^{\star}\} = \mathcal{E}(f_1^I)$ if
\begin{eqnarray}\label{eq:uniquealpha}
\alpha_i>\frac{1}{p_i^{\star}}\sum_{j > i}\alpha_j (P_j-p_j^{\star}) + \frac{\sum_{j < i} h_{ij}p_j^{\star} + \sum_{j > i} h_{ij}P_j
+n_i}{p_i^{\star}h_{i0}}
\end{eqnarray}
and
\begin{eqnarray}\label{eq:uniqueP0}
P_0 > \frac{P_i}{p_i^{\star}}\sum_{j > i}\alpha_j (P_j-p_j^{\star}) + \frac{(P_i-p_i^{\star}) ( \sum_{j < i} h_{ij}p_j^{\star} + \sum_{j > i}
h_{ij}P_j +n_i )}{p_i^{\star}h_{i0}}
\end{eqnarray}
for all $i \in \mathcal{\tilde{N}}(\mathbf{p}^{\star})$.\footnote{We define $\sum_{j \in J} x_j = 0$ if $J$ is empty. Similarly, we define $\prod_{j
\in J} x_j = 1$ if $J$ is empty.}
\end{theorem}

\begin{IEEEproof}
See \cite[Appendix~C]{XiaoParkMihaela_JSTSP}.
\end{IEEEproof}

By comparing Theorems~\ref{theorem:FirstOrderStrict_CompleteInfo} and \ref{theorem:UniqueNE_FirstOrderStrict_CompleteInfo}, we can see that the
requirements for the intervention rates and the intervention capability is higher when we impose strong sustainment. For any given power profile, the
intervention rates can be chosen sequentially to satisfy the condition \eqref{eq:uniquealpha} starting from user $N'$ down to user $1$. We can set
$\alpha_i = 0$ for all $i \notin \mathcal{\tilde{N}}(\mathbf{p}^{\star})$. Unlike Theorem~\ref{theorem:FirstOrderStrict_CompleteInfo}, the choice of
the intervention rates affects the minimum required intervention capability. For strong sustainment, the intervention capability is required to be
larger as the designer chooses larger intervention rates. A main reason for the existence of an extra NE is that the region of power profiles on
which the maximum intervention power is applied is so wide that the users cannot escape the region by unilateral deviation. Thus, a larger
intervention capability is needed to reduce the region and guarantee the uniqueness of NE.

From Theorem~\ref{theorem:FirstOrderStrict_CompleteInfo}, we obtain
\begin{eqnarray}
PB_1(\mathbf{p}^{\star}) = \max_i \frac{(P_i-p_i^{\star}) (\sum_{j\neq i} h_{ij} p_j^{\star} + n_i)}{p_i^{\star}h_{i0}}.
\end{eqnarray}
Since Theorem~\ref{theorem:UniqueNE_FirstOrderStrict_CompleteInfo} gives a sufficient condition for strong sustainment, we obtain an upper bound on
$PB_1^s(\mathbf{p}^{\star})$,
\begin{eqnarray}
\overline{PB}_1^s(\mathbf{p}^{\star}) = \sum_{i=1}^N \left[ \left(\prod_{j=1}^{i-1} \frac{P_j}{p_j^{\star}}\right) \frac{(P_i-p_i^{\star})( \sum_{j <
i} h_{ij}p_j^{\star} + \sum_{j > i} h_{ij}P_j +n_i )}{p_i^{\star}h_{i0}} \right].
\end{eqnarray}
Note that $PB_1(\mathbf{p}^{\star}) \leq \overline{PB}_1^s(\mathbf{p}^{\star})$ with equality if and only if $N' \leq 1$. Combining these results, we
can bound $PB_1^s(\mathbf{p}^{\star})$ by
\begin{eqnarray}
PB_1(\mathbf{p}^{\star}) \leq PB_1^s(\mathbf{p}^{\star}) \leq \overline{PB}_1^s(\mathbf{p}^{\star}).
\end{eqnarray}

By Theorems~\ref{theorem:FirstOrderStrict_CompleteInfo} and \ref{theorem:UniqueNE_FirstOrderStrict_CompleteInfo}, we know that first-order
intervention rules can sustain the set $\prod_i (0,P_i]$. Note that among all the efficient power profiles, those with $p_i = 0$ for some $i$ have
probability measure zero. Hence, we obtain almost the entire set of feasible payoffs by using first-order intervention rules. As argued in
Section~\ref{sec:PerformanceMetrics}, it is impossible to provide an incentive for user $i$ to choose $p_i = 0$ by intervention rules of any orders.
Thus, we actually obtain the largest set of sustainable payoffs by using first-order intervention rules. This discussion suggests that the potential
gain from using higher-order intervention rules is not from what they can sustain but how they sustain a target power profile.

\begin{remark}
An extreme intervention rule $f_e^I$ \cite{ParkMihaela_JSAC}, defined by
\begin{eqnarray}
f_e^I({\bf p}) = \left\{\begin{array}{ll}
0, & \text{if }{\bf p}=\mathbf{p}^{\star}, \\
P_0, & \text{if }{\bf p}\neq \mathbf{p}^{\star},\end{array}\right.
\end{eqnarray}
can be considered as the limiting case of first-order intervention rules as each $\alpha_i$ goes to infinity in that the area of the region
$\{\mathbf{p}: f_e^I(\mathbf{p}) \neq f_1^I(\mathbf{p}) \}$ approaches zero in the limit. With this class of intervention rules, we have
$\mathcal{E}(f_e^I) = \{\mathbf{p}^{\star}, \mathbf{P}\}$ if $P_0 \geq PB_1(\mathbf{p}^{\star})$ and $\mathcal{E}(f_e^I) = \{\mathbf{P}\}$ otherwise.
Therefore, it is impossible to construct an extreme intervention rule that strongly sustains a target power profile, except in the uninteresting case
$\mathbf{p}^{\star} = \mathbf{P}$. This motivates us to study intervention rules other than extreme intervention rules.
\end{remark}

\subsection{Dynamic Adjustment Processes}
Previously, we have derived the conditions under which a target power profile is (strongly) sustained. Now we propose a dynamic adjustment process,
in order to guide the intervention device and the users to achieve the target power profile as NE. The adjustment occurs at discrete steps, labeled
as $t=1,2,\ldots$. We allow the use of different intervention rules in different steps. Thus, the beginning of each step is triggered by the
intervention device's announcement of the intervention rule to be used in that step. Users are synchronized by the announcement of the intervention
rules. The adjustment process is based on the myopic best-response updates of the users and is described in
Algorithm~\ref{algorithm:DynamicAdjustment}.

\begin{algorithm}
\caption{A dynamic adjustment process.} \label{algorithm:DynamicAdjustment}
\begin{algorithmic}[1]
\STATE \textbf{Initialization}: $t=0$ \STATE ~~~The users choose an initial power profile $\mathbf{p}^0$. \STATE ~~~The designer announces the use of
first-order intervention rules $f^t \in \mathcal{F}_1$ with power budget $P_0$. \WHILE{${\bf p}^t\neq{\bf p}^\star$ or $f^t({\bf p}^t) \neq 0$}
\STATE $t\leftarrow t+1$ \STATE Given $\mathbf{p}^{t-1}$, the designer chooses and broadcasts the intervention rates $\alpha_i^t$ and the target
power profile $\mathbf{\tilde{p}}^t$ for the current time slot $t$.
\STATE Given $f^t({\bf p})=\left[\sum_{i=1}^N \alpha_i^t\left|p_i-\tilde{p}_i^t\right|\right]_0^{P_0}$, each user $i$ chooses a best response to
$\mathbf{p}_{-i}^{t-1}$: \STATE ~~~~~~$p_i^t \in {BR}_i(f^t,\mathbf{p}_{-i}^{t-1}) \triangleq \arg \max_{p_i \in \mathcal{P}_i}
\gamma_i(f^t,p_i,\mathbf{p}_{-i}^{t-1})$. \ENDWHILE
\end{algorithmic}
\end{algorithm}

During the adjustment process, the designer may use different intervention rules, as well as intermediate target profiles different from the final
target $\mathbf{p}^{\star}$. That is, we have $f^t \in \mathcal{F}_1(\mathbf{\tilde{p}}^t)$, where $\mathbf{\tilde{p}}^t$ is the intermediate target
power profile at step $t$. In the adjustment process, the designer chooses a sequence of intervention rules. Suppose that the designer uses an update
rule $\psi: \mathcal{P} \rightarrow \mathcal{F}_1$ to determine an intervention rule based on the power profile in the previous step. Then given an
initial power profile $\mathbf{p}^0$, an update rule yields a sequence of intervention rules and power profiles $\{(f^t,
\mathbf{p}^t)\}_{t=1}^{\infty}$.\footnote{Since the best response correspondence is non-singleton only in knife-edge cases, we focus on update rules
that yield a deterministic sequence.} We can evaluate an update rule by the following two criteria.
\begin{enumerate}
\item \emph{Convergence:} Does the induced sequence reach an equilibrium $(f,\mathbf{p}^{\star})$? If so, how many steps
are needed?
\item \emph{Minimum power budget:} How much power budget is needed to execute $\{f^t\}_{t=1}^{\infty}$, i.e.,
$\sup_t \sup_{\mathbf{p} \in \mathcal{P}} f^t(\mathbf{p})$?
\end{enumerate}

The following theorem shows that when the target power profile $\mathbf{p}^{\star}$ is close to the maximum power profile $\mathbf{P}$ and the
intervention capability $P_0$ is large, we can obtain fast convergence as well as strong sustainment.

\begin{theorem}\label{theorem:UniqueNE_FirstOrderStrict_IncompleteInfo}
For any $\mathbf{p}^{\star} \in \prod_i (0,P_i]$ such that $\sum_{i=1}^N (P_i-p_i^{\star})/P_i < 1$, there exists $(\alpha_1, \ldots, \alpha_N) \in
\mathbb{R}_+^N$ such that
\begin{eqnarray}\label{eq:strongalpha}
\alpha_i>\frac{1}{p_i^{\star}}\sum_{j \neq i}\alpha_j (P_j-p_j^{\star}) + \frac{\sum_{j \neq i} h_{ij}P_j +n_i}{p_i^{\star}h_{i0}}
\end{eqnarray}
for all $i \in \mathcal{\tilde{N}}(\mathbf{p}^{\star})$. Suppose that $f_1^s \in \mathcal{F}_{1}(\mathbf{p}^{\star})$ satisfies
\eqref{eq:strongalpha} and
\begin{eqnarray}\label{eq:strongP0}
P_0 > \frac{P_i}{p_i^{\star}}\sum_{j \neq i}\alpha_j (P_j-p_j^{\star}) + \frac{(P_i-p_i^{\star}) ( \sum_{j \neq i} h_{ij}P_j +n_i
)}{p_i^{\star}h_{i0}}
\end{eqnarray}
for all $i \in \mathcal{\tilde{N}}(\mathbf{p}^{\star})$. Then $\{\mathbf{p}^{\star}\} = \mathcal{E}(f_1^s)$. Moreover, starting from an arbitrary
initial power profile $\mathbf{p}^0 \neq \mathbf{p}^{\star}$, the adjustment process with $f^t = f_1^s$ for all $t = 1,2,\ldots$ reaches $(f_1^s,
\mathbf{p}^{\star})$ in at most two steps (and in one step if $\mathbf{p}^0 \in \prod_i [p_i^{\star},P_i]$).
\end{theorem}

\begin{IEEEproof}
See \cite[Appendix~D]{XiaoParkMihaela_JSTSP}.
\end{IEEEproof}

The minimum power budget required to execute an intervention rule described in Theorem~\ref{theorem:UniqueNE_FirstOrderStrict_IncompleteInfo} is
given by
\begin{eqnarray}
\widetilde{PB}_1^s(\mathbf{p}^{\star}) = \frac{1}{1-\sum_{i=1}^N \frac{P_i-p_i^{\star}}{P_i}} \sum_{i=1}^N \frac{(P_i-p_i^{\star}) (\sum_{j\neq i}
h_{ij} P_j + n_i)}{P_i h_{i0}}.
\end{eqnarray}
Since the requirement for $\alpha_i$ in \eqref{eq:strongalpha} is more stringent than that in \eqref{eq:uniquealpha}, we have
$\widetilde{PB}_1^s(\mathbf{p}^{\star}) \geq \overline{PB}_1^s(\mathbf{p}^{\star})$ with equality if and only if $N' \leq 1$. The difference
$\widetilde{PB}_1^s(\mathbf{p}^{\star}) - \overline{PB}_1^s(\mathbf{p}^{\star})$ can be considered as the price of fast convergence to
$\mathbf{p}^{\star}$ in terms of the minimum power budget. In addition to requiring a larger power budget,
Theorem~\ref{theorem:UniqueNE_FirstOrderStrict_IncompleteInfo} imposes a restriction on the range of target profiles. That is, the target should be
close enough to $\mathbf{P}$ for the result of Theorem~\ref{theorem:UniqueNE_FirstOrderStrict_IncompleteInfo} to hold. However, the target may not
satisfy the restriction $\sum_{i=1}^N (P_i-p_i^{\star})/P_i < 1$. In this case, the designer needs to use intermediate target power profiles that are
successively close to one another in order to guide the users to the final target. The use of intermediate target power profiles is also necessary
when the intervention device does not have a large enough power budget to strongly sustain a target power profile. In this case, the designer can use
a sequence of intervention functions to drive the users to reach the target power profile as the unique outcome. This process requires smaller power
budget than that required by strong sustainment, but may take longer time for the system to reach the target power profile.

Define the relative distance from $\mathbf{p}$ to $\mathbf{p}'$ by
\begin{eqnarray}
d(\mathbf{p},\mathbf{p}') = \sum_{i=1}^N \frac{p_i - p'_i}{p_i}.
\end{eqnarray}
Using the proofs of Theorems~\ref{theorem:UniqueNE_FirstOrderStrict_CompleteInfo} and \ref{theorem:UniqueNE_FirstOrderStrict_IncompleteInfo}, we can
show that, given $\mathbf{p}^{t-1}$, the designer can achieve the intermediate target at step $t$, i.e., $\mathbf{p}^{t} = \mathbf{\tilde{p}}^t$ only
if $\mathbf{\tilde{p}}^t$ satisfies $d(\mathbf{p}^{t-1},\mathbf{\tilde{p}}^t) < 1$. This imposes a bound on the relative distance between two
successive intermediate targets. Below we provide two different methods for the designer to generate intermediate targets. The first method, which is
summarized in Algorithm~\ref{algorithm:Adjustment_withoutP0} and analyzed in Theorem~\ref{theorem:UniqueNE_FirstOrderStrict_Convergence}, produces a
sequence of intermediate targets reaching the final target whose successive elements have a relative distance of $\delta \in (0,1)$ while requiring
the minimum power budget in each step. This method can be used in a scenario where the power constraint of the intervention device does not bind; the
designer can fix $\delta$ sufficiently close to 1, and the method will allow the system to reach the final target in the minimum number of steps. The
second method, which is summarized in Algorithm~\ref{algorithm:Adjustment_withP0} and analyzed in Theorem~\ref{theorem:T_UpperBound}, yields a
sequence of intermediate targets with the largest relative distance in each step while satisfying the power constraint. Thus, this method will allow
the manager with a limited power budget to reach the final target as fast as possible.

Now suppose that a fixed relative distance between two successive targets is given. Algorithm~\ref{algorithm:Adjustment_withoutP0} provides a method
for the designer to generate intermediate targets in the most power-budget efficient way.

\begin{algorithm}
\caption{An algorithm that generates a sequence of intermediate target power profiles with a fixed relative
distance.}\label{algorithm:Adjustment_withoutP0}
\begin{algorithmic}[1]
\REQUIRE Fix $\delta \in (0,1)$ close to 1 \STATE \textbf{Initialization ($t=1$):} Set $\mathbf{\tilde{p}}^1 = \mathbf{P}$ \COMMENT{This step can be
skipped if $\mathbf{p}^0 = \mathbf{P}$} and $\mathcal{M} = \mathcal{N}$ \STATE Set $\mu_i = p_i^\star/\tilde{p}_i^1$ for all $i\in\mathcal{N}$
\WHILE{$\sum_{i=1}^N (\tilde{p}_i^{t}-p_i^{\star})/\tilde{p}_i^{t} \geq 1$} \STATE $t\leftarrow t+1$ \WHILE{$\sum_{i=1}^N \mu_i < N - \delta$} \STATE
Choose $i^* \in \arg \max_{i \in \mathcal{M}} \frac{\sum_{j\neq i} h_{ij}\tilde{p}_j^{t-1} + n_i}{h_{i0}}$ \STATE Set $\mu_{i^*} =
\min\{1,N-\delta-\sum_{j \neq i^*} \mu_j \}$ and $\mathcal{M} \leftarrow \mathcal{M}\setminus\{i^*\}$ \ENDWHILE \STATE Set $\tilde{p}_i^t =
p_i^\star/\mu_i$ for all $i\in\mathcal{N}$ \ENDWHILE \STATE Set $\mathbf{\tilde{p}}^t = \mathbf{p}^{\star}$
\end{algorithmic}
\end{algorithm}

The following theorem shows that the designer can lead the users to the final target by using intermediate targets generated by
Algorithm~\ref{algorithm:Adjustment_withoutP0} provided that the intervention capability is sufficiently large.

\begin{theorem}\label{theorem:UniqueNE_FirstOrderStrict_Convergence}
For any $\mathbf{p}^{\star} \in \prod_i (0,P_i]$, if $\delta \geq 1 - \min_i (p_i^{\star}/P_i)$, then Algorithm~\ref{algorithm:Adjustment_withoutP0}
terminates at a finite step $T$ with $T \leq N'+1$. Let $(\mathbf{\tilde{p}}^t)_{t=1}^T$ be the sequence of power profiles generated by
Algorithm~\ref{algorithm:Adjustment_withoutP0}. Then there exists a sequence of intervention rules $(f^t)_{t=1}^T$ with $f^t \in
\mathcal{F}_1(\mathbf{\tilde{p}}^t)$ such that the adjustment process with $(f^t)_{t=1}^T$ yields $\mathbf{p}^t = \mathbf{\tilde{p}}^t$ for all $t =
1,\ldots, T$ starting from any $\mathbf{p}^0 \in \mathcal{P}$.
\end{theorem}

\begin{IEEEproof}
See \cite[Appendix~F]{XiaoParkMihaela_JSTSP}.
\end{IEEEproof}

Now we consider a scenario where the intervention capability $P_0$ should be taken into consideration while generating intermediate
targets\footnote{Note that to sustain $\mathbf{p}^\star$ as the NE, we require $P_0$ to satisfy the conditions in
Theorem~\ref{theorem:FirstOrderStrict_CompleteInfo}.}. In this scenario, in order to induce the users to follow intermediate targets during the
adjustment process, the intermediate targets $(\mathbf{\tilde{p}}^t)_{t=2}^T$ should satisfy not only
\begin{eqnarray} \label{eq:constraint1_P0}
\sum_{i=1}^{N} \frac{\tilde{p}_i^{t-1} - \tilde{p}_i^t}{\tilde{p}_i^{t-1}} < 1
\end{eqnarray}
but also
\begin{eqnarray} \label{eq:constraint2_P0}
P_0^t \triangleq \max_{i \in \mathcal{\tilde{N}}(\mathbf{\tilde{p}}^t)} \left\{ \frac{P_i}{\tilde{p}_i^{t-1}} \frac{\sum_{j=1}^N
\frac{(\tilde{p}_j^{t-1}-\tilde{p}_j^{t}) (\sum_{k\neq j} h_{jk} \tilde{p}_k^{t-1} + n_j)}{\tilde{p}_j^{t-1} h_{j0}}}{1-\sum_{j=1}^N
\frac{\tilde{p}_j^{t-1}-\tilde{p}_j^{t}}{\tilde{p}_j^{t-1}}} + \frac{(P_i - \tilde{p}_i^{t-1})(\sum_{j\neq i} h_{ij} \tilde{p}_j^{t-1} +
n_i)}{\tilde{p}_i^{t-1}h_{i0}} \right\} < P_0
\end{eqnarray}
for all $t = 2, \ldots, T$. In order to reach the final target in the minimum number of steps, we need to maximize the relative distance between
successive targets while satisfying the constraints \eqref{eq:constraint1_P0} and \eqref{eq:constraint2_P0}. Thus, the problem to obtain
$\mathbf{\tilde{p}}^{t}$ given $\mathbf{\tilde{p}}^{t-1}$ can be written as
\begin{eqnarray}
& \max_{\mathbf{\tilde{p}}^t} & \sum_{i=1}^N \frac{\tilde{p}_i^{t-1}-\tilde{p}_i^t}{\tilde{p}_i^{t-1}} \label{eq:mrdobj}\\
&s.t.& \max_{i \in \mathcal{N}(\mathbf{\tilde{p}}^{t})} \left\{\frac{P_i}{\tilde{p}_i^{t-1}} \frac{\sum_{j=1}^N
\frac{\tilde{p}_j^{t-1}-\tilde{p}_j^{t}}{\tilde{p}_j^{t-1}}b_j^{t-1}}{1-\sum_{j=1}^N \frac{\tilde{p}_j^{t-1}-\tilde{p}_j^{t}}{\tilde{p}_j^{t-1}}} +
\frac{P_i - \tilde{p}_i^{t-1}}{\tilde{p}_i^{t-1}} b_i^{t-1} \right\} \leq P_0 - \varepsilon_1 \label{eq:mrdconst1}\\
&& \sum_{i=1}^{N} \frac{\tilde{p}_i^{t-1} - \tilde{p}_i^t}{\tilde{p}_i^{t-1}} \leq 1 - \varepsilon_2 \label{eq:mrdconst2}\\
&& \mathbf{p}^{\star} \leq \mathbf{\tilde{p}}^t \leq \mathbf{\tilde{p}}^{t-1} \label{eq:mrdconst3}
\end{eqnarray}
for small $\varepsilon_1, \varepsilon_2 > 0$. Algorithm~\ref{algorithm:Adjustment_withP0} formalizes the method to generate a sequence of
intermediate target power profiles, which has maximal relative distances (MRD) between successive target power profiles given an intervention
capability $P_0$. We call the sequence generated by Algorithm~\ref{algorithm:Adjustment_withP0} the MRD sequence. Note that the major complexity in
solving the above problem is line~\ref{line:find} in Algorithm~\ref{algorithm:Adjustment_withP0}. This search on $\tilde{p}_{i^*}^t$ can be done
efficiently by bisection method, because $P_0^t$ is decreasing with $\tilde{p}_{i^*}^t$.

\begin{algorithm}
\caption{An algorithm that generates a sequence of intermediate target power profiles with maximal relative distances given an intervention
capability.}\label{algorithm:Adjustment_withP0}
\begin{algorithmic}[1]
\REQUIRE Small $\varepsilon_1 \in(0,1)$ and $\varepsilon_2 \in(0,1)$; $P_0 - \varepsilon_1 > \displaystyle\max_i \left\{ \frac{P_i-p_i^\star}{p_i^\star}\cdot\frac{\sum_{j\neq i} h_{ij}P_j+n_i}{h_{i0}}\right\}$ \STATE \textbf{Initialization ($t=1$):} Set $\mathbf{\tilde{p}}^1 = \mathbf{P}$ \COMMENT{This
step can be skipped if $\mathbf{p}^0 = \mathbf{P}$} \WHILE{$\mathbf{\tilde{p}}^t \neq {\bf p}^{\star}$} \STATE $t\leftarrow t+1$, $\mathcal{M}=\{i:
\tilde{p}_i^{t-1}>p_i^\star\},~\mathbf{\tilde{p}}^t=\mathbf{\tilde{p}}^{t-1}$ \REPEAT\label{line:repeat} \STATE $i^* = \min_{i\in\mathcal{M}}
b_i^{t-1}$, $\tilde{p}_{i^*}^t=\max\left\{p_{i^*}^\star,(N-1+\varepsilon_2-\sum_{i\neq i^*}
\tilde{p}_i^t/\tilde{p}_i^{t-1})\cdot\tilde{p}_{i^*}^{t-1}\right\}$ \STATE calculate $P_0^t$ by (\ref{eq:constraint2_P0})
\IF{$P_0^t<P_0-\varepsilon_1$} \STATE $\mathcal{M}\leftarrow\mathcal{M}\backslash\{i^*\}$ \ELSIF{$P_0^t>P_0-\varepsilon_1$} \STATE find
$\tilde{p}_{i^*}^t\in\left[\max\left\{p_{i^*}^\star,(N-1+\varepsilon_2-\sum_{i\neq i^*}
\tilde{p}_i^t/\tilde{p}_i^{t-1})\cdot\tilde{p}_{i^*}^{t-1}\right\},\tilde{p}_{i^*}^{t-1}\right]$ such that $P_0^t=P_0-\varepsilon_1$\label{line:find}
\ENDIF \UNTIL{$P_0^t=P_0-\varepsilon_1$ \textbf{or} $\mathcal{M}=\emptyset$}\label{line:until} \ENDWHILE \STATE Set $\mathbf{\tilde{p}}^t =
\mathbf{p}^{\star}$
\end{algorithmic}
\end{algorithm}

Given the power budget $P_0$, we are interested in the minimum convergence time for the dynamic adjustment process to reach the target power profile
$\mathbf{p}^\star$, defined by $T^\star(\mathbf{p}^\star)=\inf \left\{T: \mathbf{\tilde{p}}^T = \mathbf{p}^\star\right\}$, where the infimum is taken
over the set of sequences satisfying (\ref{eq:constraint1_P0}) and (\ref{eq:constraint2_P0}) starting from $\mathbf{p}^1=\mathbf{P}$. In order to obtain
an upper bound for $T^\star(\mathbf{p}^\star)$, we use an upper bound for the convergence time of a geometric sequence of $T$
intermediate target power profiles in the following form:
\begin{eqnarray}
\tilde{p}_i^t = (\eta_i)^{t-1} P_i,~\forall~i,~t=1,\ldots,T,
\end{eqnarray}
where $\eta_i = \left(p_i^\star/P_i\right)^{\frac{1}{T-1}},~i=1,\ldots,N$.

\begin{theorem}\label{theorem:T_UpperBound}
If $\mathbf{p}^0\neq\mathbf{p}^\star$, $\sum_{i=1}^N (P_i-p_i^\star)/P_i \geq 1$, and
\begin{eqnarray}\label{eqn:P0_UpperBoundT}
P_0>\left(\max_i \frac{P_i}{p_i^\star}-1\right) \max_i \frac{\sum_{j\neq i} h_{ij}p_j^\star + n_i}{h_{i0}},
\end{eqnarray}
then $T^\star(\mathbf{p}^\star)>2$ and $T^\star(\mathbf{p}^\star)$ satisfies
\begin{eqnarray}\label{eqn:ConvergenceTime_UpperBound}
\sum_{i=1}^N \left(\frac{p_i^\star}{P_i}\right)^{\frac{1}{T^\star(\mathbf{p}^\star)-2}} < N-1 + \frac{1}{C},
\end{eqnarray}
where
\begin{eqnarray}
C = \frac{P_0}{\max_i \frac{\sum_{j\neq i} h_{ij}p_j^\star + n_i}{h_{i0}} \cdot \max_i \frac{P_i}{p_i^\star}}+\frac{1}{\max_i \frac{P_i}{p_i^\star}}.
\end{eqnarray}
\end{theorem}
\begin{IEEEproof}
See \cite[Appendix~G]{XiaoParkMihaela_JSTSP}.
\end{IEEEproof}

The inequality (\ref{eqn:ConvergenceTime_UpperBound}) provides an upper bound for $T^\star(\mathbf{p}^\star)$, since the left-hand side of
(\ref{eqn:ConvergenceTime_UpperBound}) increases in $T^\star(\mathbf{p}^\star)$ and approaches $N$ as $T^\star(\mathbf{p}^\star)\rightarrow\infty$
while the right-hand side is smaller than $N$ given (\ref{eqn:P0_UpperBoundT}).
From Theorem~\ref{theorem:T_UpperBound}, we can see that the convergence time is small if the power budget $P_0$ is large, the target power profile
is close to the maximum power (i.e. $\max_i \frac{P_i}{p_i^\star}$ is small), or SINR is relatively small compared to the channel gain from the
intervention device (i.e. $\max_i \frac{\sum_{j\neq i} h_{ij}p_j^\star + n_i}{h_{i0}}$ is small).

\subsection{Relaxation of Monitoring Requirement}

The results in this section so far relies on the ability of the intervention device to estimate individual transmit powers. However, estimating
individual transmit powers requires larger monitoring overhead for the intervention device than estimating aggregate receive power. In order to study
intervention rules that can be executed with the monitoring of aggregate receive power, we consider a class of intervention rules that can be
expressed as
\begin{eqnarray} \label{eq:aggfirstorder}
f_1^A(\mathbf{p}) = \left[ \alpha_0 \left| \left(\sum_{i=1}^N  h_{0i}p_i \right) - p_A^{\star} \right| \right]_0^{P_0}
\end{eqnarray}
for some $\alpha_0 \geq 0$, $P_0 > 0$, and $p_A^{\star}$. We call an intervention rule in this class a first-order intervention rule based on
aggregate receive power or, in short, an intervention rule based on aggregate power. We call $\alpha_0$ the aggregate intervention rate, and call
$p_A^{\star}$ the target aggregate power, which is set as the aggregate receive power at the target power profile, i.e., $p_A^{\star} = \sum_{i=1}^N
h_{0i}p_i^{\star}$. We first give a necessary and sufficient condition for an intervention rule based on aggregate power to sustain a target power
profile.

\begin{theorem}\label{theorem:FirstOrderAggregate}
For any $\mathbf{p}^{\star} \in \prod_i (0,P_i]$, $\mathbf{p}^{\star} \in \mathcal{E}(f_1^A)$ if and only if
\begin{eqnarray} \label{eq:aggalphai}
\alpha_0 \geq \max_{i \in \mathcal{\tilde{N}}(\mathbf{p}^{\star})} \frac{\sum_{j\neq i} h_{ij} p_j^{\star} + n_i}{h_{0i} p_i^{\star} h_{i0}}
\end{eqnarray}
and
\begin{eqnarray} \label{eq:aggP0firstorder}
P_0 \geq \max_{i \in \mathcal{\tilde{N}}(\mathbf{p}^{\star})} \frac{(P_i-p_i^{\star}) (\sum_{j\neq i} h_{ij} p_j^{\star} + n_i)}{p_i^{\star}h_{i0}}.
\end{eqnarray}
\end{theorem}

\begin{IEEEproof}
See \cite[Appendix~H]{XiaoParkMihaela_JSTSP}.
\end{IEEEproof}

The minimum intervention capability required to sustain a target profile is not affected by using aggregate receive power instead of individual
transmit powers. However, the aggregate intervention rate should be chosen high enough to prevent a deviation of any user, whereas with the
monitoring of individual transmit powers the intervention rates can be chosen individually for each user. This suggests that strong sustainment is
more difficult with intervention rules based on aggregate power. For example, $\mathbf{P}$ is also sustained by $f_1^A$ if $P_0 \leq \alpha_0 \sum_{j
\neq i} (h_{0j} P_j - p_j^{\star})$ for all $i \in \mathcal{\tilde{N}}(\mathbf{p}^{\star})$, which is weaker than the corresponding condition in the
case of intervention rules based on individual powers, $P_0 \leq \sum_{j \neq i} \alpha_j (P_j - p_j^{\star})$ for all $i \in
\mathcal{\tilde{N}}(\mathbf{p}^{\star})$. With the monitoring of individual powers, a deviation of each user can be detected and punished. This leads
to the property that the best response of user $i$ is almost always either $p_i^{\star}$ or $P_i$ under first-order intervention rules based on
individual powers. This implies that a power profile sustained by a first-order intervention rule based on individual powers almost always belongs to
the set $\prod_i \{p_i^{\star}, P_i\}$. In contrast, with the monitoring of aggregate power, only an aggregate deviation can be detected. This yields
a possibility that an intervention rule based on aggregate power sustains a power profile that is different from the target but yields the same
aggregate power. This possibility makes the problem of coordination failure more worrisome because if the users are given only the target aggregate
power $p_A^{\star}$ they may not know which power profile to select among those that yield the aggregate power $p_A^{\star}$.\footnote{A way to
overcome this problem is to broadcast the target power profile $\mathbf{p}^{\star}$ to the users in order to make $\mathbf{p}^{\star}$ as a focal
point \cite{fudenberg}.} The problem arising from the increased degree of non-uniqueness can be considered as the cost of reduced monitoring
overhead. To state the result formally, let $\alpha_0^i = (\sum_{j\neq i} h_{ij} p_j^{\star} + n_i)/(h_{0i} p_i^{\star} h_{i0})$ and $P_0^i =
(P_i-p_i^{\star}) (\sum_{j\neq i} h_{ij} p_j^{\star} + n_i) / (p_i^{\star}h_{i0})$ for all $i \in \mathcal{\tilde{N}}(\mathbf{p}^{\star})$. Also, let
$\bar{\alpha}_0 = \max_{i \in \mathcal{\tilde{N}}(\mathbf{p}^{\star})} \alpha_0^i$ and $\bar{P}_0 = \max_{i \in
\mathcal{\tilde{N}}(\mathbf{p}^{\star})} P_0^i$.

\begin{theorem}\label{theorem:FirstOrderAggregate_multiple}
Suppose that, for $\mathbf{p}^{\star} \in \prod_i (0,P_i]$, there exist $i, j \in \mathcal{\tilde{N}}(\mathbf{p}^{\star})$ such that (i)
$\bar{\alpha}_0 = \alpha_0^i > \alpha_0^j$ or $\alpha_0^i, \alpha_0^j < \bar{\alpha}_0$, and (ii) $\bar{P}_0 = P_0^i > P_0^j$ or $P_0^i, P_0^j <
\bar{P}_0$. Then for any $f_1^A$ such that $\mathbf{p}^{\star} \in \mathcal{E}(f_1^A)$ and for any $\epsilon > 0$, there exists $\mathbf{\tilde{p}}
\neq \mathbf{p}^{\star}$ such that $\mathbf{\tilde{p}} \in \mathcal{E}(f_1^A)$, $\sum_{i=1}^N  h_{0i}p_i^{\star} = \sum_{i=1}^N  h_{0i}\tilde{p}_i$,
and $| \mathbf{\tilde{p}} - \mathbf{p}^{\star}| < \epsilon$.
\end{theorem}

\begin{IEEEproof}
See \cite[Appendix~I]{XiaoParkMihaela_JSTSP}.
\end{IEEEproof}

Theorem~\ref{theorem:FirstOrderAggregate_multiple} provides a sufficient condition under which the strong sustainment of a given target power profile
is impossible with intervention rules based on aggregate power. We argue that the sufficient condition is mild. First, note that, for almost all
$\mathbf{p}^{\star} \in \prod_i (0,P_i]$, $\alpha_0^i$'s and $P_0^i$'s can be ordered strictly. With strict ordering of $\alpha_0^i$'s and $P_0^i$'s,
we can always find a pair of users $i, j \in \mathcal{\tilde{N}}(\mathbf{p}^{\star})$ satisfying the condition in
Theorem~\ref{theorem:FirstOrderAggregate_multiple} if there are at least three users in $\mathcal{\tilde{N}}(\mathbf{p}^{\star})$. That is, strong
sustainment is generically impossible with intervention rules based on aggregate power when $|\mathcal{\tilde{N}}(\mathbf{p}^{\star})| \geq 3$.

\section{Implementation Issues} \label{sec:implementation}
In this section, we discuss some implementation issues. First, we provide algorithms for the intervention device to estimate the system parameters
and individual transmit powers without user cooperation. Then we compare the communication overhead of the intervention scheme with that of other
incentive schemes.

\subsection{Estimation of System Parameters and Individual Transmit Powers}
As we have seen from previous results (e.g.
Theorems~\ref{theorem:FirstOrderStrict_CompleteInfo}--\ref{theorem:UniqueNE_FirstOrderStrict_IncompleteInfo}), in order to determine the intervention
rates and the power budget requirement, the designer needs to know the normalized cross channel gains $\{\frac{h_{ij}}{h_{i0}}\}_{j=1,j\neq i}^N,
\forall i\in\mathcal{N}$, the normalized noise powers $\frac{n_i}{h_{i0}}, \forall i\in\mathcal{N}$, the maximum power levels $P_i, \forall
i\in\mathcal{N}$, and the target power profile $\mathbf{p}^\star$. We propose a method to estimate the normalized cross channel gains, normalized
noise powers, and the maximum power levels without user cooperation. Based on the above parameters, the designer can determine the target power
profile $\mathbf{p}^\star$ by solving \eqref{eq:desprob}. In addition, we propose a method to estimate the individual transmit powers without user
cooperation.

\subsubsection{Normalized cross channel gains and noise powers}
The designer estimates the normalized cross channel gains and the normalized noise powers by adjusting the intervention rules and observing the
reaction of the users. First, the designer broadcast the intervention capability $P_0$ and a temporary target power profile
$\mathbf{\tilde{p}}<\mathbf{P}$. Then it makes $N$ rounds of measurements at $N$ different locations. We assume that during the measurements, the
users always choose the power levels that maximize their SINRs given the intervention rule. Thus, we exclude the strategic behavior of the users to
influence the measurements in their favor. We also assume that the intervention device can move its receiver to $N$ different locations, or it has
$N$ receivers located at different locations.

\begin{algorithm}
\caption{The $n$th round of measurement performed by the intervention device.}\label{algorithm:Measurement_Round_n}
\begin{algorithmic}[1]
\REQUIRE Error tolerance $\varepsilon>0$ \STATE \textbf{Initialization:} Broadcast $\alpha_j=0,~\forall j\in\mathcal{N}$ \label{algorithm:broadcast1}
\FOR{$\mathrm{index}=0$ to $N-1$} \STATE Set $i = (n+\mathrm{index)~mod}~N$, $\underline{\alpha}_i=0$, $\bar{\alpha}_i=0$, $\alpha_i$ any positive
value (preferably large) \STATE Measure the aggregate receive power $\bar{R}_i$ and set the current aggregate receive power $R_i=\bar{R}_i$
\WHILE{$\bar{\alpha}_i=0$ \textbf{or} $\bar{\alpha}_i-\underline{\alpha}_i>\varepsilon$} \IF{$\bar{\alpha}_i=0$ \textbf{and} $R_i=\bar{R}_i$} \STATE
$\alpha_i\leftarrow 2\cdot\alpha_i$ \ELSIF{$R_i=\bar{R}_i$} \STATE $\underline{\alpha}_i\leftarrow\alpha_i$,
$\alpha_i\leftarrow(\underline{\alpha}_i+\bar{\alpha}_i)/2$ \ELSE \STATE $\bar{\alpha}_i\leftarrow\alpha_i$,
$\alpha_i\leftarrow(\underline{\alpha}_i+\bar{\alpha}_i)/2$ \ENDIF \STATE Broadcast the index $i$ and the new $\alpha_i$,
\label{algorithm:broadcast2} and measure the current aggregate receive power $R_i$ \IF{$R_i<\bar{R}_i$} \STATE Set $\underline{R}_i=R_i$ \ENDIF
\ENDWHILE \ENDFOR
\end{algorithmic}
\end{algorithm}

In round $n$, the designer adjusts the intervention rates one by one, starting from $\alpha_n$ to $\alpha_N$, then from $\alpha_1$ to $\alpha_{n-1}$.
All measurements are made at the receiver at location $n$. When adjusting $\alpha_i$, the aim is to find $\alpha_i^*$, the minimum intervention rate
at which user $i$'s best response is $\tilde{p}_i$. We can calculate $\alpha_i^*$ as
\begin{eqnarray}\label{eqn:Measurement_InterventionRate}
\alpha_i^* = \left\{\begin{array}{ll}\frac{\sum_{j=1}^{i-1} h_{ij}\tilde{p}_j + \sum_{j=i+1}^{n-1} h_{ij}P_j + \sum_{j=n}^N h_{ij}\tilde{p}_j + n_i}{h_{i0}}, & i<n \\
\frac{\sum_{j=1}^{n-1} h_{ij}P_j + \sum_{j=n}^{i-1} h_{ij}\tilde{p}_j + \sum_{j=i+1}^N h_{ij}P_j + n_i}{h_{i0}}, & i\geq n \end{array}\right..
\end{eqnarray}
The designer tunes $\alpha_i$ to find $\alpha_i^*$ according to the change of the aggregate receive power. When $\alpha_i>\alpha_i^*$, user $i$'s
best response is $P_i$, and the aggregate receive power at location $n$ is
\begin{eqnarray}
\bar{R}_i^n = \left\{\begin{array}{ll} \sum_{j=1}^{i-1} h_{0j}^n \tilde{p}_j + \sum_{j=i}^{n-1} h_{0j}^n P_j + \sum_{j=n}^N h_{0j}^n \tilde{p}_j + n_0, & i<n \\
\sum_{j=1}^{n-1} h_{0j}^n P_j + \sum_{j=n}^{i-1} h_{0j}^n \tilde{p}_j + \sum_{j=i}^N h_{0j}^n P_j + n_0, & i\geq n \end{array}\right.,
\end{eqnarray}
where $h_{0j}^n$ is the channel gain from user $j$'s transmitter to the intervention device's receiver at location $n$. When $\alpha_i>\alpha_i^*$,
user $i$'s best response is the target power profile $\tilde{p}_i$, and the aggregate receive power decreases to
\begin{eqnarray}
\underline{R}_i^n = \left\{\begin{array}{ll} \sum_{j=1}^{i} h_{0j}^n \tilde{p}_j + \sum_{j=i+1}^{n-1} h_{0j}^n P_j + \sum_{j=n}^N h_{0j}^n \tilde{p}_j + n_0, & i<n \\
\sum_{j=1}^{n-1} h_{0j}^n P_j + \sum_{j=n}^{i} h_{0j}^n \tilde{p}_j + \sum_{j=i+1}^N h_{0j}^n P_j + n_0, & i\geq n \end{array}\right..
\end{eqnarray}
During the measurement, the designer maintains an upper bound $\bar{\alpha}_i$, at which the aggregate receive power is $\underline{R}_i$, and a
lower bound $\underline{\alpha}_i$, at which the aggregate receive power is $\bar{R}_i$. By bisection methods, an estimate of $\alpha_i^*$, denoted
by $m_{ni}$, is obtained within the error tolerance $\varepsilon$. The $n$th--round measurement is summarized in
Algorithm~\ref{algorithm:Measurement_Round_n}.

Round $n$ returns a set of measurements $\{\bar{R}_i^n,\underline{R}_i^n\}_{i=1}^N$, from which we obtain maximum power levels $\{P_i\}_{i=1}^N$.
Note that $\bar{R}_i^n-\underline{R}_i^n=h_{0i}^n\cdot(P_i-\tilde{p}_i)$ for all $i,n\in\mathcal{N}$. Thus, we have
\begin{eqnarray}
h_{0i}^n = h_{0i}^1\cdot\frac{\bar{R}_i^n-\underline{R}_i^n}{\bar{R}_i^1-\underline{R}_i^1},~\forall i,n\in\mathcal{N}.
\end{eqnarray}
Since $\underline{R}_{n-1}^n = \sum_{j=1}^N h_{0j}^n \tilde{p}_j$ ($\underline{R}_0^1=\underline{R}_N^1$ when $n=1$), using the above relationship
between $h_{0i}^n$ and $h_{0i}^1$, we have the following $N$ linear equations
\begin{eqnarray}
\sum_{j=1}^N h_{0j}^1\cdot\left(\frac{\bar{R}_j^n-\underline{R}_j^n}{\bar{R}_j^1-\underline{R}_j^1} \tilde{p}_j\right) =
\underline{R}_{n-1}^n,~n=1,\ldots,N,
\end{eqnarray}
which we can solve for $\{h_{0j}^1\}_{j=1}^N$. Given $\{h_{0j}^1\}_{j=1}^N$, we can calculate $\{P_i\}_{i=1}^N$ by using
$\bar{R}_i^1-\underline{R}_i^1=h_{0i}^1\cdot(P_i-\tilde{p}_i), \forall i\in\mathcal{N}$.

Round $n$ also returns another set of measurements ${\bf m}_n = \left[m_{n1},~\ldots,~m_{ni},~\ldots,~m_{nN}\right]$, where we assume
$m_{ni}=\alpha_i^*$. Given the measurements $\{\mathbf{m}_n\}_{n=1}^N$, we can obtain the normalized cross channel gains and the normalized noise
powers. Specifically, from ${\bf m}_n$ and ${\bf m}_{n+1}, \forall~n<N$, we have
\begin{eqnarray}
m_{n+1,i}-m_{ni} = \frac{h_{in}P_n-h_{in}\tilde{p}_n}{h_{i0}},~\forall~i\neq n,
\end{eqnarray}
from which we can get $\frac{h_{in}}{h_{i0}}=(m_{n+1,i}-m_{ni})/(P_n-\tilde{p}_n), \forall~i\neq n$.

To sum up, we can get $\{\frac{h_{in}}{h_{i0}}\}_{i\neq n}$ according to ${\bf m}_n$ and ${\bf m}_{n+1}$ for all $n<N$, and get
$\{\frac{h_{iN}}{h_{i0}}\}_{i\neq N}$ according to ${\bf m}_N$ and ${\bf m}_1$. Now that we know all the normalized channel gains, we can get the
normalized noise powers $\frac{n_i}{h_{i0}}$ from \eqref{eqn:Measurement_InterventionRate}.

\subsubsection{Individual Transmit Powers}
The byproduct of the above estimation is the channel gains from the users to the intervention device at $N$ different locations:
$\{h_{0i}^n\}_{i=1}^N, n=1,\ldots,N$. At any time, the intervention device can measure the aggregate received power at all the $N$ locations
\begin{eqnarray}
\sum_{j=1}^N h_{0j}^n p_j + n_0^n,
\end{eqnarray}
where $n_0^n$ is the noise power known to the intervention device. Since the designer knows the values of $N$ different linear combinations of the
$N$ individual transmit powers, it can solve the group of $N$ linear equations to obtain the individual transmit powers. The complexity of this
operation is of the order $N^3$.

\subsection{Comparison of Communication Overhead}
Now we compare the communication overhead of the intervention scheme with that of other frameworks, including network utility maximization
\cite{HuangBerry_JSAC06}\cite{HandeChiang_ToN08}, game theoretic control based on taxation \cite{AlpcanBasar_CDMA}--\cite{HuangBerry_06_Auction}, and
mechanism design \cite{SharmaTeneketzis_ToN}\cite{SharmaTeneketzis_GameTheory}. Before the comparison, we would like to emphasize that the
intervention scheme works for selfish users, who have no incentive to provide any information to anyone else. As shown in
Table~\ref{table:ComparisonCommunicationOverhead}, intervention requires no information flow from the users to the designer. On the contrary, the
other works assume that the users are obedient to exchange information with the designer or among each other according to some prescribed rules.

The communication overhead is measured by the total amount of information flow before the system reaches the desired operating point. Specifically,
the amount of information flow is measured by the number of real numbers broadcast by the users and the intervention device, ignoring quantization
and coding. The communication overhead can be further categorized into the communication overhead on the users and that on the designer. We summarize
the comparison in Table~\ref{table:ComparisonCommunicationOverhead}. In each framework, we select representative algorithms to calculate the
communication overhead. Hence, the numbers in the table are not precise for all the algorithms, but are correct in terms of the order.

\begin{table*}
\renewcommand{\arraystretch}{1.1}
\caption{Comparison of Communication Overhead of Different Frameworks} \label{table:ComparisonCommunicationOverhead} \centering
\begin{tabularx}{\textwidth}{|c|c|c|c|}
\hline
Framework & Users & The Designer & Overall \\
\hline
Intervention & $0$ & $2N^2\log_2(1/\varepsilon)+2N\cdot T^\star(\mathbf{p}^\star)$ & $2N^2\log_2(1/\varepsilon)+2N\cdot T^\star(\mathbf{p}^\star)$ \\
\hline
Network utility maximization & $N$ (\cite{HuangBerry_JSAC06}) or $0$ (\cite{HandeChiang_ToN08}) & $0$ (\cite{HuangBerry_JSAC06}) or $N$ (\cite{HandeChiang_ToN08}) & \multirow{2}{*}{$N$ in each step} \\ \cite{HuangBerry_JSAC06}\cite{HandeChiang_ToN08} & in each step & in each step & \\
\hline
Game theoretic control & \multirow{2}{*}{$N$ in each step} & \multirow{2}{*}{$N$ in each step} & \multirow{2}{*}{$2N$ in each step} \\ based on taxation \cite{AlpcanBasar_CDMA}--\cite{HuangBerry_06_Auction} & & & \\
\hline
Mechanism design \cite{SharmaTeneketzis_ToN}\cite{SharmaTeneketzis_GameTheory} & $N^2$ in each step & $N$ in each step & $N^2+N$ in each step \\
\hline
\end{tabularx}
\end{table*}

We can see from Table~\ref{table:ComparisonCommunicationOverhead} that in intervention, users have zero communication overhead. In other words, the
intervention device does not rely on the users to provide information. While in other frameworks, users may be required to broadcast some
information. Hence, intervention is more suitable when users are selfish and unwilling to provide information truthfully. Another advantage of
intervention is that its communication overhead can be bounded analytically. While in other frameworks, the convergence speed to the desired
operating point is not guaranteed. The detailed analysis is as follows.

\subsubsection{Intervention}
The communication overhead of intervention is comprised of two parts. The first part is the overhead of estimating normalized cross channel gains and
noise powers. The temporary target power profile $\mathbf{\tilde{p}}$ and the intervention capability $P_0$ is broadcast at the beginning, which we
omit here. There are $N$ rounds of measurement. In each round, the intervention device needs to broadcast the initial intervention rates
(line~\ref{algorithm:broadcast1} in Algorithm~\ref{algorithm:Measurement_Round_n}), and the indices and the values of the intervention rates it is
adjusting (line~\ref{algorithm:broadcast2} in Algorithm~\ref{algorithm:Measurement_Round_n}). Since it uses bisection methods to adjust the
intervention rate, the number of adjustments is $\log_2(1/\varepsilon)$ for each intervention rate, where $\varepsilon$ is the error tolerance in
estimating intervention rates in Algorithm~\ref{algorithm:Measurement_Round_n}. In sum, the communication overhead in estimating system parameters is
$N(2N\log_2(1/\varepsilon)+1)\approx 2N^2\log_2(1/\varepsilon)$ for all intervention rates.

The second part is the communication overhead during the convergence to NE. In general, this overhead depends on the target power profile and the
intervention capability. If the target power profile is close to the maximum power level ($\sum_{i=1}^N (P_i-p_i^\star)/Pi<1$) and the intervention
capability is large (Theorem~\ref{theorem:UniqueNE_FirstOrderStrict_IncompleteInfo}), the convergence is in one step. If the target power profile is
not close enough or the intervention capability is limited, we need to use the dynamic adjustment process. The convergence is in $N'+1$ steps if the
intervention capability is large (Theorem~\ref{theorem:UniqueNE_FirstOrderStrict_Convergence}), and is in $T^\star(\mathbf{p}^\star)$ steps if the
intervention capability is limited (Theorem~\ref{theorem:T_UpperBound}). In each step, the intermediate target power profile and the intervention
rates are broadcast. In summary, the worst-case communication overhead during the convergence to NE is $2N\cdot T^\star(\mathbf{p}^\star)$.

\subsubsection{Other frameworks}
The communication overhead of other frameworks is the total information flow during the convergence to the desired operating point. Unlike
intervention, the number of steps during the convergence is not bounded. Hence, the communication overhead could be arbitrarily large.

If users cooperate with the designer to maximize the assigned utility function, as in \cite{HuangBerry_JSAC06}\cite{HandeChiang_ToN08} and
\cite{AlpcanBasar_CDMA}--\cite{HuangBerry_06_Auction}, the information flow in one step can be less than that in intervention. In network utility
maximization, either each user or the base station will broadcast at least $N$ signals for all the users. In \cite{HuangBerry_JSAC06}, each user
broadcast its own interference price. In \cite{HandeChiang_ToN08}, the base station broadcast ``load'' and ``spillage''. In game theoretic control
based on taxation \cite{AlpcanBasar_CDMA}--\cite{HuangBerry_06_Auction}, the base station broadcast $N$ prices, usually different for different
users. To obtain the optimal pricing, users report back some information, such as their payoffs or interferences, to adjust prices.

In mechanism design \cite{SharmaTeneketzis_ToN}\cite{SharmaTeneketzis_GameTheory}, similar to intervention, the designer does not know the utility
function of each user. In this case, the information flow in one step is of the same order as that in intervention. Specifically, each user broadcast
its own version of optimal power allocation vector in each step, and the designer broadcast the reference power allocation vector. Hence, the amount
of information flow is $N^2+N$ in each step.

\section{Simulation Results} \label{sec:Simulation}
\begin{figure}
\centering
\includegraphics[width =4.0in]{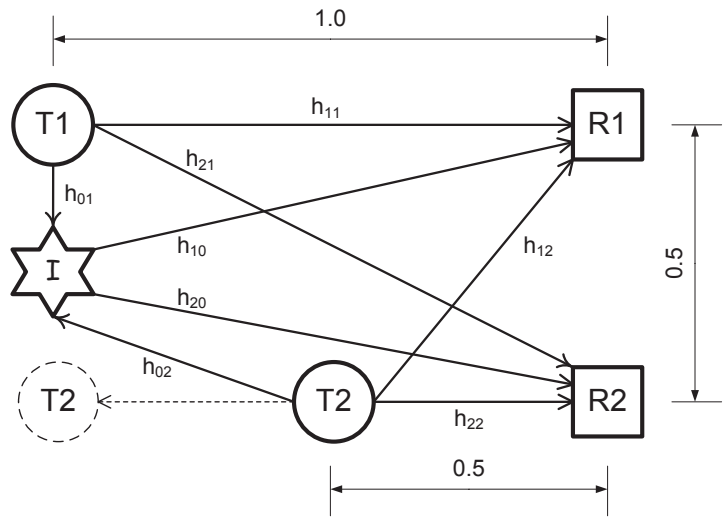}
\caption{An example single-hop wireless ad-hoc network with two users.} \label{AdHoc_TwoUser}
\end{figure}
We consider a two-user network shown in Fig. \ref{AdHoc_TwoUser}. User 2's transmitter is near to user 1's receiver, causing significant interference
to user 1. The distance from user 1's transmitter to its receiver is normalized to $1$. Originally, the distance from user 2's transmitter to its
receiver is $0.5$. The vertical distance between the two users' transmitters and that between the two users' receivers are both $0.5$. Without
specific notice, we assume that the positions of the transmitters and receivers of both users remain the same. In the simulation of
Fig.~\ref{fig:SocialWelfare_2user}, we let user 2's transmitter move away from its receiver as shown by the dashed left arrow, resulting in less
interference to user 1. The channel gain $h$ is reciprocal to the distance $d$ with the path loss exponent $a$, namely $h=d^{-a}$. We assume an
indoor environment where $a=3$ \cite{Rappaport}. The noise powers at the receivers of both users are $0.2$. The power budgets of both users are 10.

\subsection{Performance Improvement By Intervention}
\begin{figure}
\centering
\includegraphics[width =4.0in]{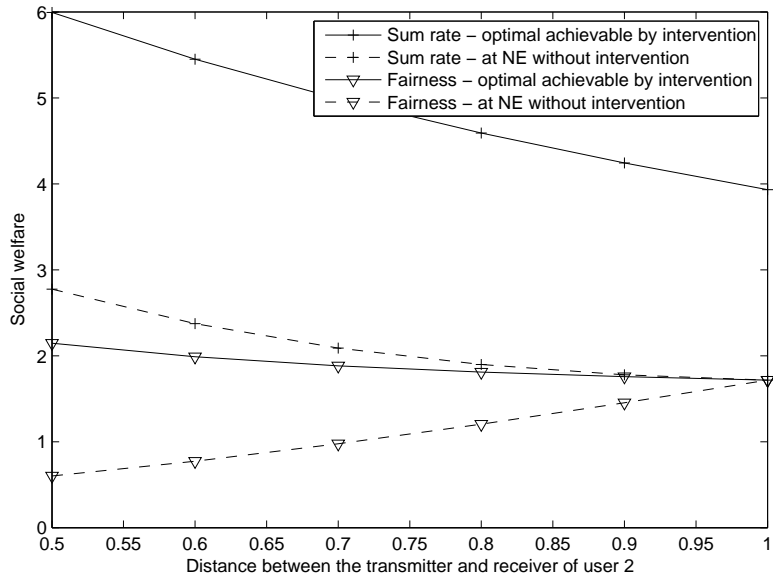}
\caption{The optimal social welfare achievable by intervention and the social welfare at NE without intervention, when user 2's transmitter moves
away from its receiver.} \label{fig:SocialWelfare_2user}
\end{figure}
Now we examine the performance improvement by using intervention mechanisms. We let user 2's transmitter moves away from its receiver. In Fig.
\ref{fig:SocialWelfare_2user}, we show the performance achieved by intervention and that at the NE without intervention, under two criteria for
social welfare. The sum rate is define by
\begin{eqnarray}
\log\left(1+\gamma_1\right)+\log\left(1+\gamma_2\right),
\end{eqnarray}
and the fairness is defined by the ``max-min'' fairness \cite[pp.~392]{Chiang_FoundationTrend08}\cite{HandeChiang_ToN08}\cite{SchubertBoche}
\begin{eqnarray}
\min\left\{\log\left(1+\gamma_1\right),~\log\left(1+\gamma_2\right)\right\}.
\end{eqnarray}

As we can see from Fig. \ref{fig:SocialWelfare_2user}, the sum rate achievable by intervention doubles that at the NE without intervention in all the
cases. The fairness achievable by intervention is much larger than that at the NE without intervention in most cases. When the distance from user 2's
transmitter to its receiver is 1.0, the network is symmetric. Only at this point is the NE without intervention optimal in terms of fairness.

\subsection{Minimum Power Budget}
\begin{figure*}
\centering{\subfloat[]{\includegraphics[width=3.2in]{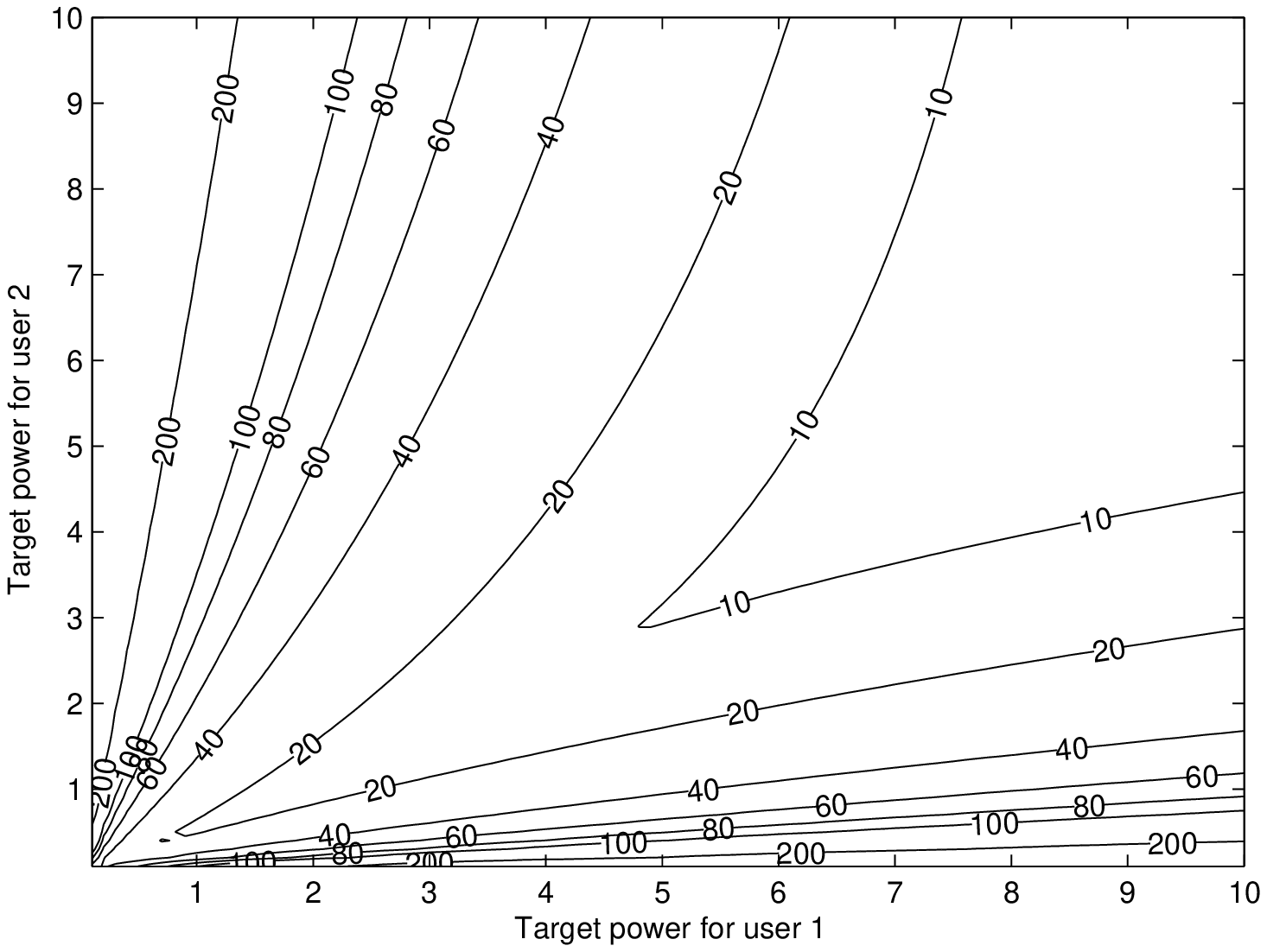}\label{fig:MinPowerContour_2user_gs1_NE_Pos1}}
\subfloat[]{\includegraphics[width=3.2in]{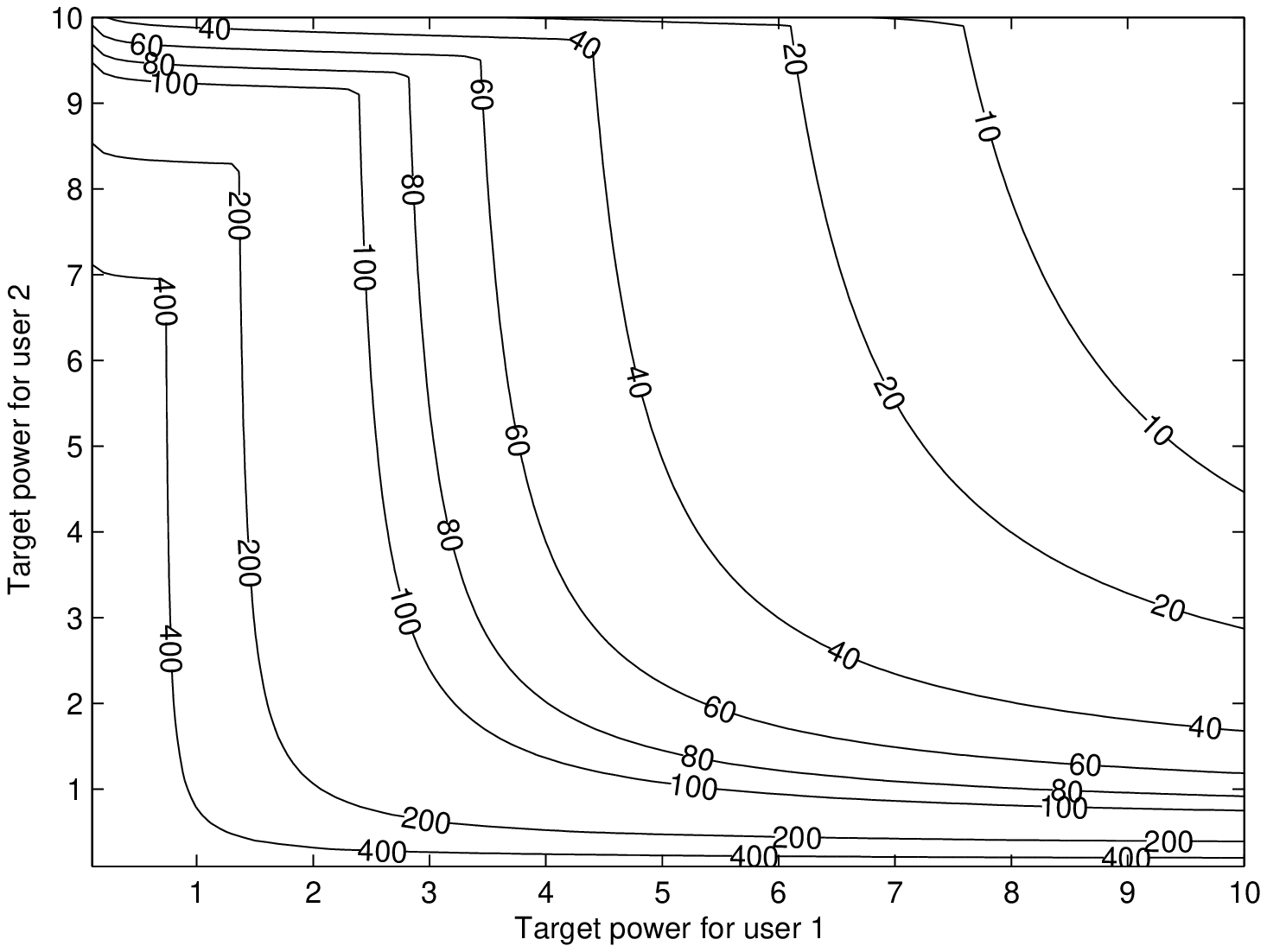}\label{fig:MinPowerContour_2user_gs1_SufficientNecessary_Pos1}}\\
\subfloat[]{\includegraphics[width=3.2in]{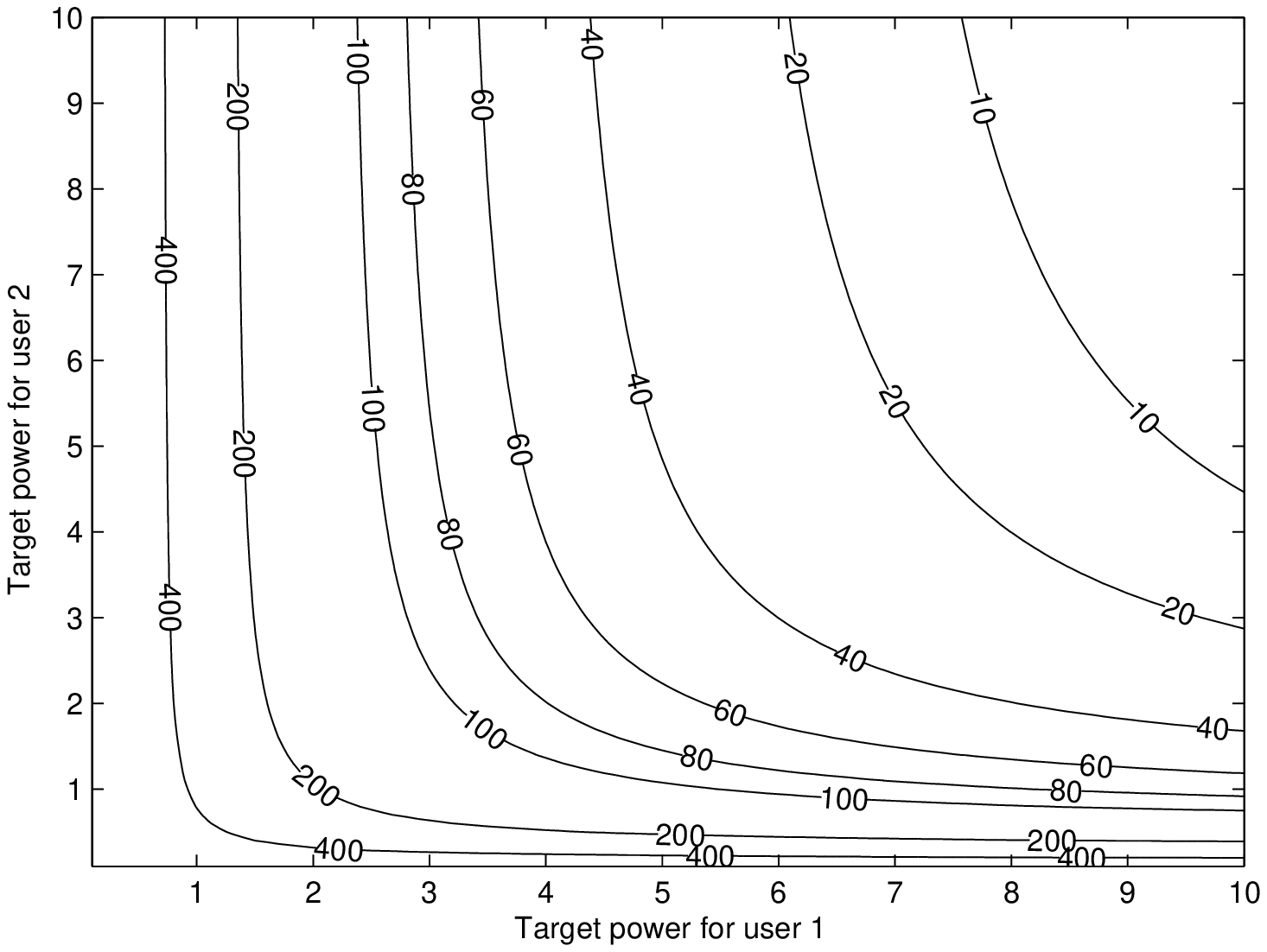}\label{fig:MinPowerContour_2user_gs1_CompleteInfo_Pos1}}
\subfloat[]{\includegraphics[width=3.2in]{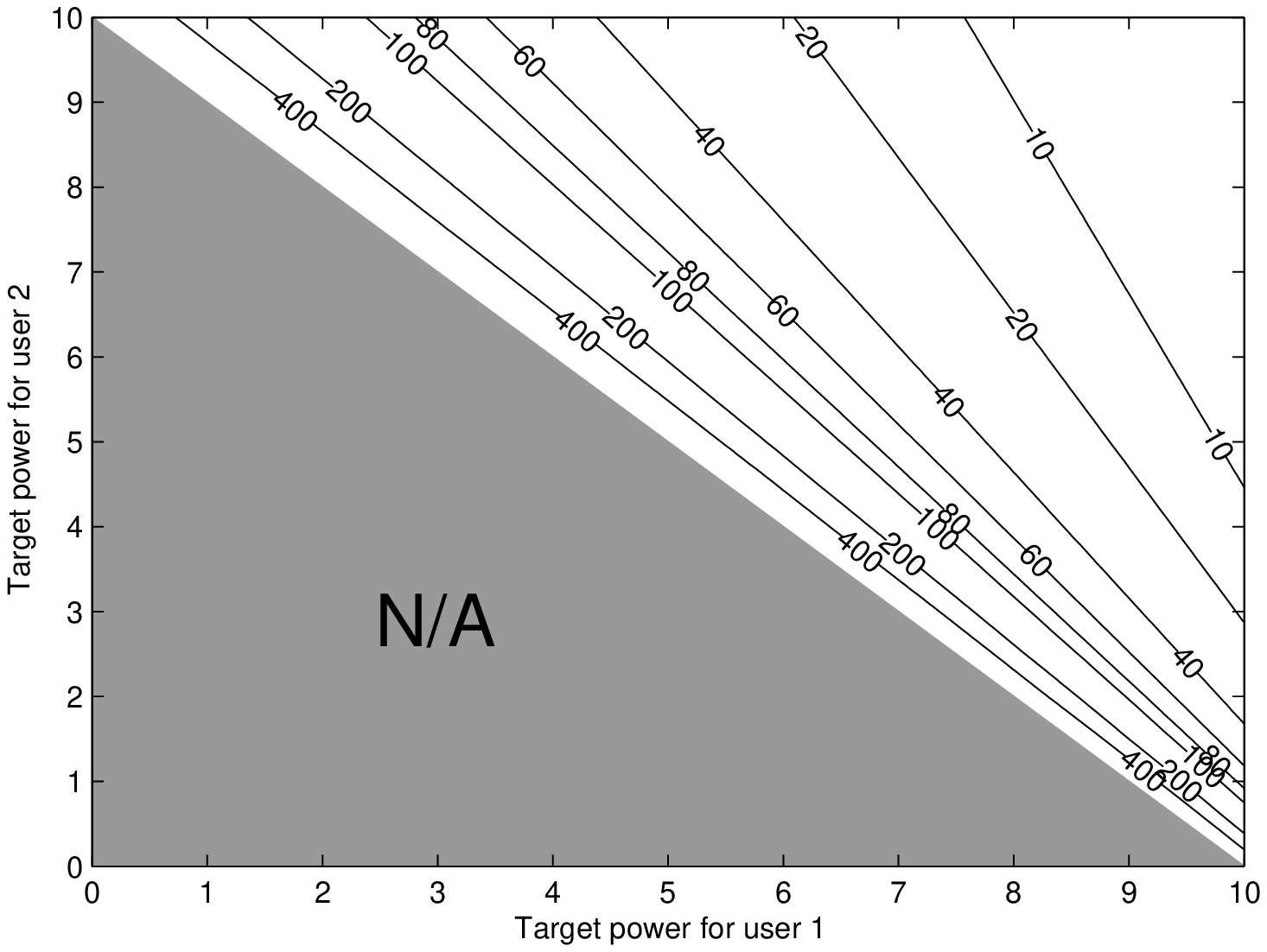}\label{fig:MinPowerContour_2user_gs1_IncompleteInfo_Pos1}}}
\caption{Contour of the minimum power budget of first-order intervention under different target power profiles. (a): minimum power budget to sustain
a target power profile obtained by Theorem~\ref{theorem:FirstOrderStrict_CompleteInfo} and Theorem~\ref{theorem:FirstOrderAggregate}; (b): minimum
power budget to strongly sustain a target power profile obtained by simulation; (c): upper bound on the minimum power budget to strongly sustain a
target power profile obtained by Theorem~\ref{theorem:UniqueNE_FirstOrderStrict_CompleteInfo}; (d): upper bound on the minimum power budget for
strong sustainment and fast convergence obtained by Theorem~\ref{theorem:UniqueNE_FirstOrderStrict_IncompleteInfo}.}
\label{fig:MinPowerContour_2user_Pos1}
\end{figure*}

Now we show the power budget requirement for different intervention rules. In Fig.~\ref{fig:MinPowerContour_2user_Pos1}, we show the contour of the
minimum power budget for different intervention rules under different target power profiles.
Fig.~\ref{fig:MinPowerContour_2user_Pos1}\subref{fig:MinPowerContour_2user_gs1_NE_Pos1} shows minimum power budget to sustain a target power profile
using first-order intervention based on individual transmit powers obtained by Theorem~\ref{theorem:FirstOrderStrict_CompleteInfo} and that using
first-order intervention based on aggregate receive power obtained by Theorem~\ref{theorem:FirstOrderAggregate}. Since the power budget requirements
are the same for these two intervention rules, we show them in the same figure.
Fig.~\ref{fig:MinPowerContour_2user_Pos1}\subref{fig:MinPowerContour_2user_gs1_SufficientNecessary_Pos1} shows the minimum power budget to strongly
sustain a target power profile obtained by simulation. As we expect, the power budget requirement for strong sustainment is higher.
Fig.~\ref{fig:MinPowerContour_2user_Pos1}\subref{fig:MinPowerContour_2user_gs1_CompleteInfo_Pos1} shows the upper bound on the minimum power budget
to strongly sustain a target power profile obtained by Theorem~\ref{theorem:UniqueNE_FirstOrderStrict_CompleteInfo}. We can see that the result in
Theorem~\ref{theorem:UniqueNE_FirstOrderStrict_CompleteInfo} serves as a good upper bound. Finally,
Fig.~\ref{fig:MinPowerContour_2user_Pos1}\subref{fig:MinPowerContour_2user_gs1_IncompleteInfo_Pos1} shows the upper bound on the minimum power budget
for strong sustainment and fast convergence obtained by Theorem~\ref{theorem:UniqueNE_FirstOrderStrict_IncompleteInfo}. In this case, the system
reaches NE in at most two time slots. To achieve this fast convergence, the intervention device needs a much higher power budget. In addition, not
all the target power profiles can be sustained. The target power profiles that cannot be sustained lie in the shadow area in the figure.

\subsection{Power Budget and Convergence Time Tradeoff In Dynamic Adjustment Process}
\begin{figure*}
\centering{\subfloat[]{\includegraphics[width =3.2in]{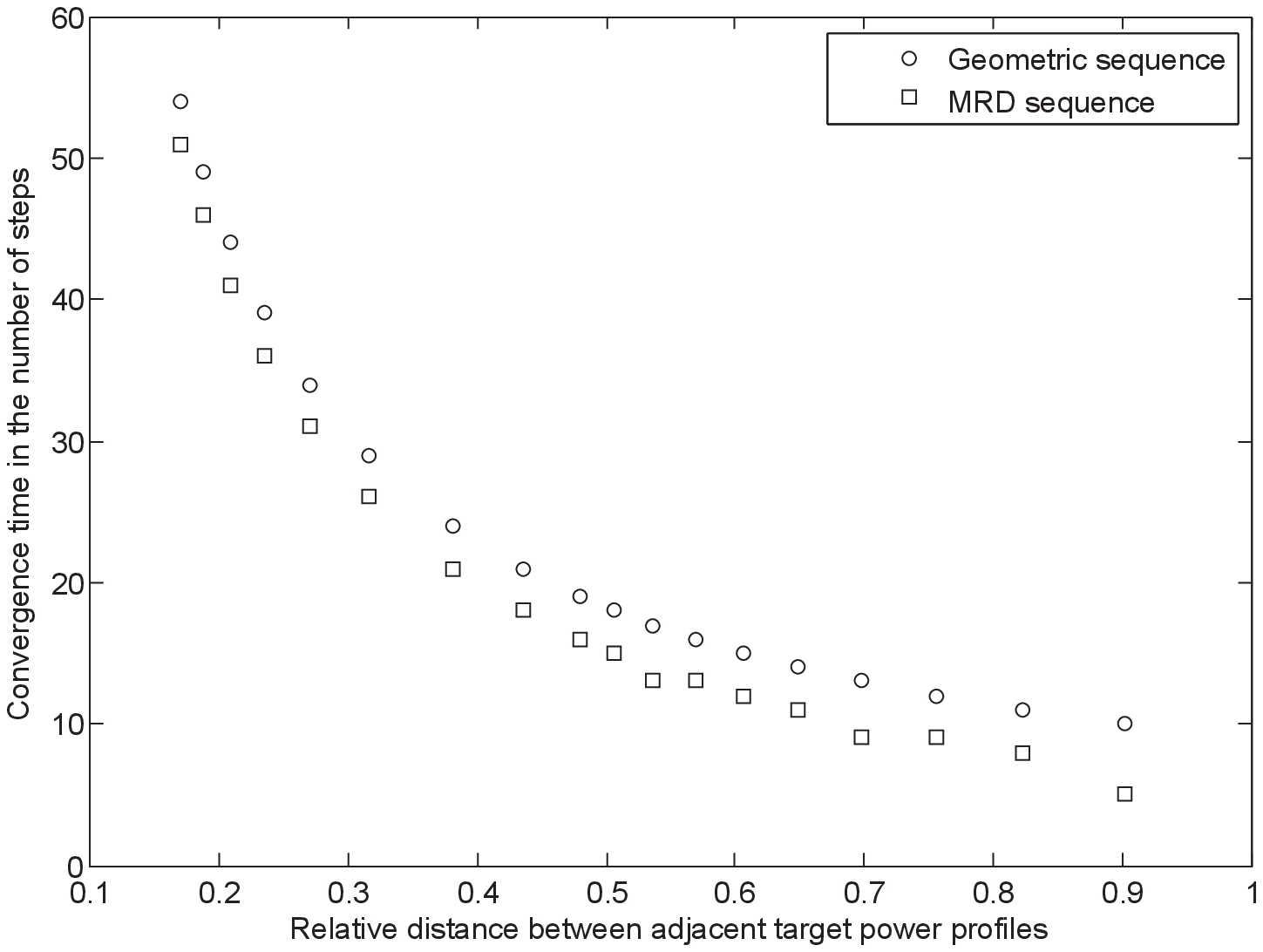}\label{fig:Adjustment_5user_GivenRD_T}}
\subfloat[]{\includegraphics[width =3.2in]{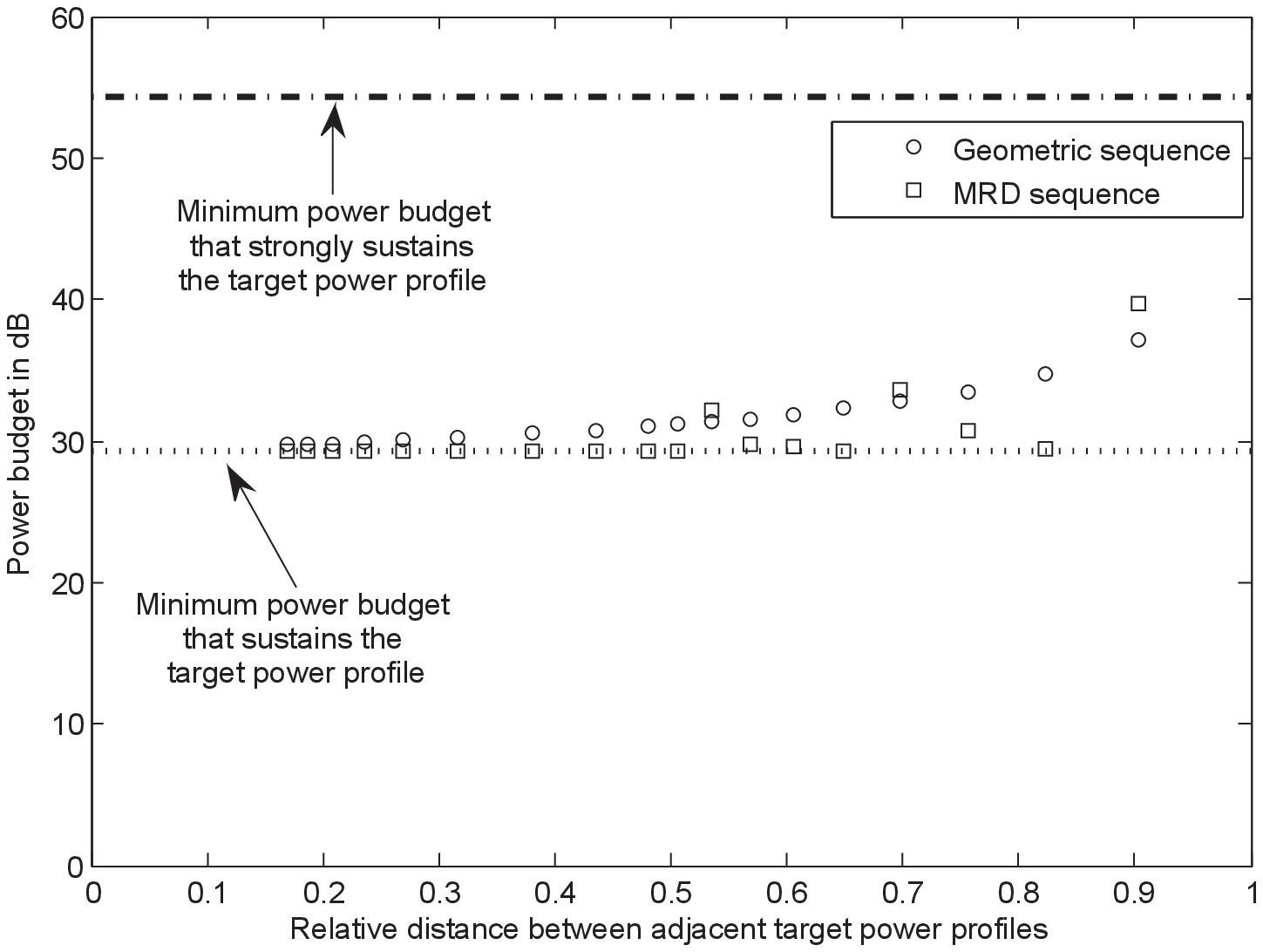}\label{fig:Adjustment_5user_GivenRD_P0}}} \caption{Given the relative
distance between adjacent target power profiles, the convergence time and the power budget requirement of different sequences of intermediate target
power profiles in a five-user network. The relative distance between the maximum power profile and the target power profile is $\sum_{i=1}^5
\frac{P_i-p_i^\star}{P_i}=3.6>1$. (a): convergence time; (b): power budget requirement.} \label{fig:Adjustment_5user_GivenRD}
\end{figure*}

\begin{figure}
\centering
\includegraphics[width=4.0in]{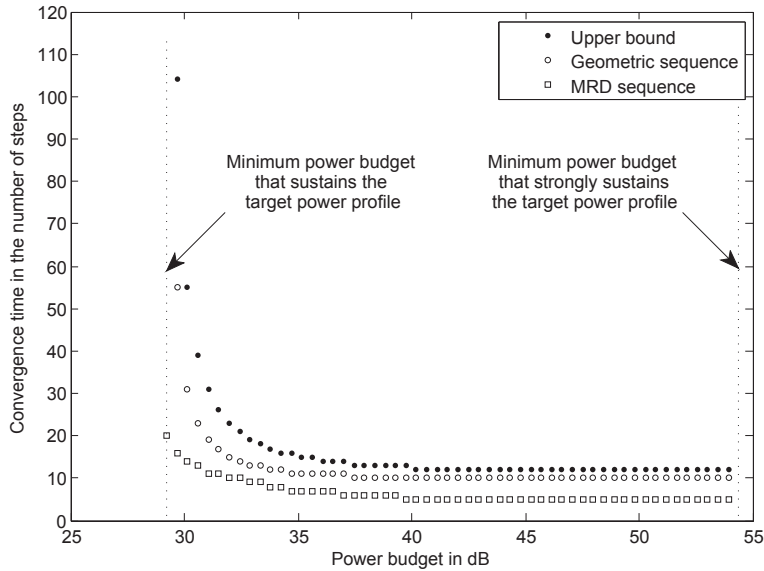}
\caption{Given power budget requirements, the convergence time of different sequences of intermediate target power profiles in a five-user network.
The relative distance between the maximum power profile and the target power profile is $\sum_{i=1}^5 \frac{P_i-p_i^\star}{P_i}=3.6>1$.}
\label{fig:Adjustment_5user_GivenP0}
\end{figure}

Now we study the tradeoff between the power budget and the convergence time in the dynamic adjustment process. Here the convergence time is measured
as the number of steps in the adjustment process. To better illustrate the tradeoff, we use a five-user network in the simulation for
Fig.~\ref{fig:Adjustment_5user_GivenRD} and Fig.~\ref{fig:Adjustment_5user_GivenP0}. The channel gains and noise powers used in the simulation are
one realization of the random variables. Since different realizations result in similar tradeoff, we only show the results for one realization. The
target power profile is $p_1^\star=P_1$ and $p_i^\star=0.1P_i$ for $i>1$. Since the relative distance between the maximum power profile and the
target power profile is $\sum_{i=1}^5 \frac{P_i-p_i^\star}{P_i}=3.6>1$, we cannot reach the target power profile from the maximum power profile
directly using Theorem~\ref{theorem:UniqueNE_FirstOrderStrict_IncompleteInfo}. Instead, we need a sequence of intermediate target power profile
before we reach the final target power profile.

First, suppose that there is no power budget requirement. Without power budget limit, our goal is to reach the final target power profile in as few
time slots as possible. Since it is not easy to construct a sequence of intermediate target power profiles given a desired convergence time, we
construct the sequence according to the desired relative distance between adjacent intermediate target power profiles, which is an indicator for the
convergence time. In Fig.~\ref{fig:Adjustment_5user_GivenRD}, we show the convergence time and the power budget requirement of the MRD sequence
generated by Algorithm~\ref{algorithm:Adjustment_withoutP0} and the geometric sequence under different relative distances. We can see from
Fig~\ref{fig:Adjustment_5user_GivenRD}\subref{fig:Adjustment_5user_GivenRD_T} that a larger relative distance results in a faster convergence for
both sequences. Thus, we can use the relative distance, a metric amenable for the construction of the sequence, to control the convergence speed of
the adjustment process. In particular, when the relative distance is $\delta=0.9=1-\min_i \{p_i^\star/P_i\}$, the MRD sequence converges in
$N^\prime+1=5$ steps as predicted by Theorem~\ref{theorem:UniqueNE_FirstOrderStrict_Convergence}. In
Fig~\ref{fig:Adjustment_5user_GivenRD}\subref{fig:Adjustment_5user_GivenRD_P0}, we can see that for both sequences, the power budget requirement is
decreasing with the relative distances in most cases. The power budget is lower for the MRD sequence. For both sequences, it requires much less power
by the dynamic adjustment process than by the strong sustainment condition in Theorem~\ref{theorem:UniqueNE_FirstOrderStrict_CompleteInfo}.

Second, suppose that there is a power budget requirement. Given different power budget requirements, we show the convergence time of MRD and
geometric sequences and the upper bound on the convergence time in Fig.~\ref{fig:Adjustment_5user_GivenP0}. We can see from the figure that under
most power budget requirements, the convergence time of MRD sequence is roughly half of that of the naive geometric sequence. When the power budget
is small (near to the minimum power budget that sustains the target power profile), the fast convergence of MRD sequence is even more significant
compared to that of the geometric sequence. When the power budget approaches the minimum requirement that strongly sustains the target power profile,
the convergence time of MRD sequence is 5, which is half of that of the geometric sequence.

\section{Conclusion}\label{sec:conclusion}
In this paper, we proposed incentive schemes based on intervention for power control in single-hop wireless ad hoc networks with selfish users. We
formulated a game-theoretic model of power control with an intervention device and proposed design criteria that desirable intervention rules should
satisfy. Focusing on a simple class of intervention rules called first-order intervention rules, we provided requirements for intervention rules to
sustain a target power profile when the intervention device estimates individual transmit powers or aggregate receive power. We also proposed dynamic
adjustment processes to guide users to a target power profile through intermediate targets. We discussed implementation issues and presented
simulation results. To the best of our knowledge, this work is the first to investigate intervention schemes in a power control scenario. For future
research, we can apply intervention schemes to different power control scenarios, for example, a scenario where users can allocate their power
budgets across multiple channels and a scenario where users care about their energy consumption as well as their data rates.

%

\begin{IEEEbiography}{Yuanzhang Xiao}
received the B.E. and M.E. degree in electrical engineering from Tsinghua University, Beijing, China, in 2006 and 2009, respectively. He is currently
pursuing the Ph.D. degree in the Electrical Engineering Department at University of California, Los Angeles. His research interests include game
theory, optimization, communication networks, and network economics.
\end{IEEEbiography}

\begin{IEEEbiography}{Jaeok Park}
received the B.A. degree in economics from Yonsei University, Seoul, Korea, in 2003, and the M.A. and Ph.D. degrees in economics from the University
of California, Los Angeles, in 2005 and 2009, respectively.

He is currently an Assistant Professor in the School of Economics at Yonsei University, Seoul, Korea. From 2009 to 2011, he was a Postdoctoral
Scholar in the Electrical Engineering Department at the University of California, Los Angeles. From 2006 to 2008, he served in the Republic of Korea
Army. His research interests include game theory, mechanism design, network economics, and wireless communication.
\end{IEEEbiography}


\begin{IEEEbiographynophoto}{Mihaela van der Schaar}
is Chancellor's Professor of Electrical Engineering at University of California, Los Angeles. She is an IEEE Fellow, a Distinguished Lecturer of the
Communications Society for 2011-2012 and the Editor in Chief of IEEE Transactions on Multimedia. She received an NSF CAREER Award (2004), the Best
Paper Award from IEEE Transactions on Circuits and Systems for Video Technology (2005), the Okawa Foundation Award (2006), the IBM Faculty Award
(2005, 2007, 2008), the Most Cited Paper Award from EURASIP: Image Communications Journal (2006), the Gamenets Conference Best Paper Award (2011) and
the 2011 IEEE Circuits and Systems Society Darlington Award Best Paper Award. She holds 33 granted US patents. For more information about her
research visit: http://medianetlab.ee.ucla.edu/
\end{IEEEbiographynophoto}


\begin{thebibliography}{10}
\bibitem{Chiang_FoundationTrend08}
M.~Chiang, P.~Hande, T.~Lan, and C.~W. Tan, ``Power control in wireless cellular networks,'' \emph{Foundations and Trends in Networking}, vol. 2, no.
4, pp. 381--533, Apr. 2008.

\bibitem{Yates95}
R.~D. Yates, ``A framework for uplink power control in cellular radio systems,'' \emph{IEEE J. Sel. Areas Commun.}, vol. 13, no. 7, pp. 1341--1347,
Sep. 1995.

\bibitem{Altman03}
E.~Altman and Z.~Altman, ``S-modular games and power control in wireless networks,'' \emph{IEEE Trans. Autom. Control}, vol. 48, no. 5, pp. 839--842,
May 2003.

\bibitem{HuangBerry_JSAC06}
J.~Huang, R.~A. Berry, and M.~L. Honig, ``Distributed interference compensation for wireless networks,'' \emph{IEEE J. Sel. Areas Commun.}, vol. 24,
no. 5, pp. 1074--1084, May 2006.

\bibitem{HandeChiang_ToN08}
P.~Hande, S.~Rangan, M.~Chiang, and X.~Wu, ``Distributed uplink power control for optimal SIR assignment in cellular data networks,'' \emph{IEEE/ACM
Trans. Netw.}, vol. 16, no. 6, pp. 1420--1433, Dec. 2008.

\bibitem{Altman_Survey06}
E.~Altman, T.~Boulogne, R.~El-Azouzi, T.~Jimenez, and L.~Wynter, ``A survey on networking games in telecommunications,'' \emph{Computers and
Operations Research}, vol. 33, no. 2, pp. 286--311, Feb. 2006.

\bibitem{LarssonJorswieck09}
E.~G. Larsson, E.~A. Jorswieck, J.~Lindblom, and R.~Mochaourab, ``Game theory and the flat-fading gaussian interference channel,'' \emph{IEEE Signal
Process. Mag.}, vol. 26, no. 5, pp. 18--27, Sep. 2009.

\bibitem{AlpcanBasar_CDMA}
T.~Alpcan, T.~Basar, R.~Srikant, and E.~Altman, ``CDMA uplink power control as a noncooperative game,'' \emph{Wireless Networks}, vol. 8, pp.
659--670, 2002.

\bibitem{SaraydarMandayam_TCOM02}
C.~U. Saraydar, N.~B. Mandayam, and D.~J. Goodman, ``Efficient power control via pricing in wireless data networks,'' \emph{IEEE Trans. Commun.},
vol. 50, no. 2, pp. 291--303, Feb. 2002.

\bibitem{XiaoShroff_ToN}
M.~Xiao, N.~B. Shroff, and E.~K. P. Chong, ``A utility-based power control scheme in wireless cellular systems,'' \emph{IEEE/ACM Trans. Netw.}, vol.
11, no. 2, pp. 210--221, Apr. 2003.

\bibitem{ScutariPalomar_ICASSP06}
G.~Scutari, S.~Barbarossa, and D.~P. Palomar, ``Potential games: A framework for vector power control problems with coupled constraints,'' in
\emph{Proceeding of ICASSP 2006}, pp. 241--244, 2006.

\bibitem{CandoganOzdaglar_INFOCOM10}
U.~O. Candogan, I.~Menache, A.~Ozdaglar, and P.~A. Parrilo, ``Near-optimal power control in wireless networks: a potential game approach,'' in
\emph{Proceedings of IEEE INFOCOM 2010}, pp. 1--9, 2010.

\bibitem{HuangBerry_06_Auction}
J.~Huang, R.~A. Berry, and M.~L. Honig, ``Auction-based spectrum sharing,'' \emph{Mobile Networks and Applications}, vol. 11, pp. 405--418, 2006.

\bibitem{SharmaTeneketzis_ToN}
S.~Sharma and D.~Teneketzis, ``An externalities-based decentralized optimal power allocation algorithm for wireless networks,'' \emph{IEEE/ACM Trans.
Netw.}, vol. 17, no. 6, pp. 1819--1831, Dec. 2009.

\bibitem{SharmaTeneketzis_GameTheory}
S.~Sharma and D.~Teneketzis, ``A game-theoretic approach to decentralized optimal power allocation for cellular networks,'' \emph{Telecommunication
Systems}, pp. 1--16, 2010.

\bibitem{WuLiu09}
Y.~Wu, B.~Wang, K.~J. R. Liu, and T.~C. Clancy, ``Repeated open spectrum sharing game with cheat-proof strategies,'' \emph{IEEE Trans. Wireless
Commun.}, vol. 8, no. 4, pp. 1922--1933, Apr. 2009.

\bibitem{LongZhang_JSAC07}
C.~Long, Q.~Zhang, B.~Li, H.~Yang, and X.~Guan, ``Non-cooperative power control for wireless ad hoc networks with repeated games,'' \emph{IEEE J.
Sel. Areas Commun.}, vol. 25, no. 6, pp. 1101--1112, Aug. 2007.

\bibitem{mailath} G. Mailath and L. Samuelson, \emph{Repeated Games
and Reputations: Long-run Relationships}. Oxford, U.K.: Oxford Univ. Press, 2006.

\bibitem{ParkMihaela_EURASIP}
J.~Park and M.~van der Schaar, ``Stackelberg contention games in multiuser networks,'' \emph{EURASIP Journal on Advances in Signal Processing}, vol.
2009, pp. 1--15, 2009.

\bibitem{ParkMihaela_Gamenets}
J.~Park and M.~van der Schaar, ``Designing incentive schemes based on intervention: The case of imperfect monitoring,'' in {\it Proceedings of
GameNets 2011}. Available: http://medianetlab.ee.ucla.edu/papers/gamenets2011\_jaeok.pdf.


\bibitem{ParkMihaela_JSAC}
J.~Park and M.~van der Schaar, ``The theory of intervention games for resource sharing in wireless communications,'' Accepted by \emph{IEEE J. Sel.
Areas Commun.} Available: http://arxiv.org/abs/1101.3052.

\bibitem{GaiKrishnamachari_Infocom}
Y. Gai, H. Liu, and B. Krishnamachari, ``A packet dropping-based incentive mechanism for M/M/1 queues with selfish users'', in \emph{Proceedings of
IEEE INFOCOM 2011}, pp. 2687--2695, 2011.

\bibitem{XiaoParkMihaela_JSTSP}
Y.~Xiao, J.~Park, and M.~van der Schaar, ``Intervention in Power Control Games With Selfish Users,'' \emph{Tech. Rep.} Available:
http://arxiv.org/abs/1104.1227v1.

\bibitem{fudenberg} D. Fudenberg and J. Tirole, \emph{Game Theory}.
Cambridge, MA: MIT Press, 1991.

\bibitem{Rappaport}
T.~S. Rappaport, \emph{Wireless Communications: Principles and Practice}. Upper Saddle River, N.J.: Prentice Hall PTR, 2002.

\bibitem{SchubertBoche}
M.~Schubert and H.~Boche, ``Solution of the multiuser downlink beamforming problem with individual SINR constraints,'' \emph{IEEE Trans. Veh.
Technol.}, vol. 53, no. 1, pp. 18--28, Jan. 2004.

\bibitem{HornJohnson}
R.~A. Horn and C.~R. Johnson, \emph{Matrix Analysis}. Cambridge, U.K.: Cambridge Univ. Press, 1985.

\end{thebibliography}
\end{document}